\documentclass[12pt]{article}
\usepackage{epsfig, amsmath, amssymb}
\usepackage{hyperref}
\usepackage{multicol}
\usepackage{graphicx,psfrag}
\usepackage{tikz}
\usetikzlibrary{decorations.pathmorphing,decorations.markings} % for wiggly lines & arrows
\usepackage{physics}
\usepackage{xcolor}
\usepackage{cite}
\usepackage{mathtools,slashed}
\usepackage{textpos}
\usepackage[utf8]{inputenc}
\usepackage[T1]{fontenc}
\usepackage{helvet}
\usepackage[]{hyperref}

\usepackage{slashed,verbatim,subfigure}
\usepackage{appendix}
\catcode`@11
\def\seceqaa{\@addtoreset{equation}{section}
	\def\theequation{A\arabic{equation}}}
\def\seceqbb{\@addtoreset{equation}{section}
	\def\theequation{B\arabic{equation}}}
	\def\seceqcc{\@addtoreset{equation}{section}
	\def\theequation{C\arabic{equation}}}
\usepackage{physics}
\newcommand{\nn}{\nonumber}
\newcommand{\be}{\begin{equation}}
\newcommand{\ee}{\end{equation}}
\newcommand{\ben}{\begin{equation}}
\newcommand{\een}{\end{equation}}
\newcommand{\bea}{\begin{eqnarray}}
\newcommand{\eea}{\end{eqnarray}}
\newcommand{\bA}{\begin{array}}
\newcommand{\eA}{\end{array}}
\newcommand{\bc}{\begin{center}}
\newcommand{\ec}{\end{center}}
\newcommand{\al}{\alpha}

\newcommand{\ra}{\rightarrow}
\newcommand{\del}{\partial}

\newcommand{\ie}{{\it i.e.}}

\usepackage{multirow}
\newcommand{\ba}{\begin{eqnarray}}
\newcommand{\ea}{\end{eqnarray}}

\usepackage{graphicx,psfrag}
\usepackage{xcolor}
\numberwithin{equation}{section}

\begin{document}

%\ifeprint
%\fi

\begin{titlepage}
%\vspace{30mm}

%\begin{flushright}
%
%\end{flushright}

\bc

%\hfill  {TIFR/TH/09-12} \\
\hfill % {\tt arXiv:0909.4731 [hep-th]} 
\\         [25mm]
%X\vfill

{\Large \bf $AdS/CFT$ to $dS/CFT$: Some Recent Developments}
% Lectures on AdS/CFT correspondence (or ``holography'')} 
%\begin{flushright}\vspace{-4mm}{\footnotesize K. Narayan}\end{flushright}
\vspace{10mm}

{\renewcommand{\thefootnote}{$\dagger$}
\large {\bf Gopal Yadav}\footnote{\small Email: gopal12896@gmail.com, gopalyadav@cmi.ac.in}}
 \\
\vspace{3mm}
{\small \it Chennai Mathematical Institute, \\}
{\small \it H1 SIPCOT IT Park, Siruseri 603103, India.\\}

\ec
%\medskip
\vspace{15mm}

\begin{abstract}
These lecture notes aim to provide a pedagogical introduction to the AdS/CFT correspondence and its extensions to spacetimes with positive (de Sitter spacetime) and zero (flat spacetime) cosmological constant. We begin by explaining the physical motivation for holography and the significance of the AdS/CFT correspondence. We then review the basic ingredients of conformal field theory (CFT) and anti de Sitter (AdS) spacetime required to formulate the duality. Building on these foundations, we discuss the formulation of the AdS/CFT correspondence and discuss several consistency checks that support it. We conclude with a brief discussion of holography in de Sitter and flat spacetimes.

%In this lecture notes, we provide a brief overview of the AdS/CFT correspondence and its generalization to spacetimes with positive (de Sitter spacetime) and zero (flat spacetime) cosmological constant. For this purpose, we start with the motivation that why and how AdS/CFT correspondence is important. We then introduce the basics of conformal field theory (CFT) and anti de Sitter (AdS) spacetime. Using these ingredients, we discuss the details of AdS/CFT correspondence with few consistency checks. We end this lecture notes with discussion on how holography works in de Sitter and flat spacetimes.

 %These lecture notes will provide an overview of AdS/CFT correspondence. We will start with why AdS/CFT is needed and then go through basics of CFT and AdS spacetime. Using these results, we will provide the details of AdS/CFT correspondence followed by explicit examples and generalization of this duality to other spacetimes.
\end{abstract}
\end{titlepage}

\tableofcontents

\section{Introduction and motivation}
These lecture notes is based on the pedagogical  lectures given at ST$^4$ in 2025 (\url{https://st4physics.wixsite.com/2025}). We had a total of six lectures, each 1.5 hours. These lectures aim to provide a partial overview of AdS/CFT correspondence (in general ``holography'') to the graduate students and early career postdocs unfamiliar with this research area. For more detailed understanding of holography, see also \cite{Bigatti:1999dp, Klebanov:2000me, DHoker:2002nbb, Maldacena:2003nj, Polchinski:2010hw, Natsuume:2014sfa, Ammon:2015wua, DeWolfe:2018dkl} apart from the references mentioned in this lecture notes.
So, keeping this in mind, we will discuss some basics of conformal field theory (CFT) in sec. \ref{CFT}, anti de Sitter (AdS) spacetime in sec. \ref{AdSmd} and then discuss the AdS/CFT correspondence, followed by Maldacena's conjecture, which requires some basic knowledge of string theory and $D$ branes, so we discuss these also before the lecture on Maldacena's conjecture in sec. \ref{AdS/CFT}. To make the readers more comfortable, we discuss three consistency checks of AdS/CFT correspondence in sec. \ref{C-checks}: although there are many nowadays. We encourage the readers to look at more literature. The AdS/CFT correspondence has been generalized to de Sitter space and flat spacetime; hence, in the last lecture, we briefly discussed how holography works in these backgrounds. Prerequisites for these lectures are quantum field theory and general theory of relativity. We discuss the basics before every advanced topic so that readers can follow these lectures easily.

%We are living in a universe that has four types of interactions: strong, weak, electromagnetic, and gravitational. Three of them are well studied using quantum field theory (QFT) in a framework known as the ``Standard Model of Particle Physics''. We can't describe gravity using quantum field theory. So the natural question one can ask is how to quantize gravity? The candidate theory where we can see quantization of gravity and the unification of all four interactions in nature is ``string theory'' where the fundamental object is a string. See \cite{Green:2012oqa, Polchinski:1998rq, Polchinski:1998rr} for the follow up discussion in this paragraph for more details. We can find scalar fields, gauge fields, and gravitons from the quantization of strings. First, we were able to discuss only bosons from string theory, which is known as ``bosonic string theory''. Later fermions were also included in string theory using supersymmetry (supersymmetry relates bosons with fermions and vice-versa), which leads to ``superstring theory''. There are five versions of superstring theory: type $IIB$, type $IIA$, type $I$, Heterotic $SO(32)$ and Heterotic $E_8 \times E_8$. Later, in 1995, Witten found that these five versions of superstring theory could be unified in a single framework known as ``M theory'' \cite{Witten:1995ex}.

We can study the weak coupling regime of quantum field theory using perturbation theory. In the strongly coupled regime, we can't apply perturbation theory. The question one can ask is how to explore the strongly coupled regime of QFTs? Remarkably, in 1997, Maldacena proposed a duality between $N=4$ SYM theory in four dimensions and type $IIB$ supergravity in $AdS_5 \times S^5$ background \cite{Maldacena:1997re}. Here SYM is understood as ``supersymmetric Yang-Mills''\footnote{In Electrodynamics, we have an abelian gauge field, whereas in Yang-Mills theories we have non-abelian gauge fields.}. The duality proposed by Maldacena relates the strongly coupled regime of QFTs and the weakly coupled regime of gravitational theories in general. There is a mapping between the parameters of these two different theories. Therefore, we can compute the relevant quantities of QFTs in the bulk theory using AdS/CFT duality and then translate our results to QFTs using the aforementioned mapping. The AdS/CFT duality has been tested in several examples, see \cite{Casalderrey-Solana:2011dxg, Hartnoll:2016apf, Berges:2020fwq, Czajka:2018bod}, and it is important because
\begin{itemize}
\item AdS/CFT duality is needed to explore the strongly coupled regime of quantum field theories.
\item AdS/CFT paradigm can resolve the information paradox \cite{Hawking:1975vcx, Hawking:1976ra}.
\item It has applications in condensed matter physics (AdS/CMT) \cite{Hartnoll:2009sz, Sachdev:2010ch}, black holes \cite{Almheiri:2020cfm, Raju:2020smc, Mahajan:2025gfh}, QCD \cite{Mia:2009wj, Kim:2012ey, Dhuria:2013tca, Yadav:2020tyo}, open effective field theory \cite{Loganayagam:2020eue}, flat space \cite{Ball:2019atb,Cheung:2016iub,Nguyen:2022zgs,Donnay:2022hkf,Laddha:2022nmj}, cosmological singularities \cite{Das:2006dz, Awad:2007fj, Engelhardt:2014mea}, higher spin systems \cite{Sleight:2017krf}
, etc.
\item One can construct multiverse models using holography \cite{Garriga:2008ks, Dil:2020fvf, Yadav:2023qfg, Aguilar-Gutierrez:2023zoi, Yadav:2024ray}.
\end{itemize}
We provided a few cases where AdS/CFT is useful, but there are also more. After Maldacena's seminal work \cite{Maldacena:1997re}, we have studied this duality a lot for AdS spacetime. But we are living in a universe that is different from AdS. We are living in de Sitter space, which has a positive cosmological constant. How holography works in de Sitter space is not well understood yet. We will discuss various versions of dS holography in sec. \ref{dSholography} and then make some comments on the holography in flat space in sec. \ref{fsholography}. With all these discussions in these lectures, we end our lectures and encourage the readers to go through the literature to learn more about holography.

We used the following partial list of references \cite{Qualls:2015qjb, Nawata:2022ywk, Kaplan, Nastase, Bayona:2005nq, Penedones:2016voo, Hubeny:2014bla, GSW} for these lectures. These are not the full list of references that we used for these lectures. At some places, we used specific papers to discuss certain topics that we have mentioned at those places.

\section{Conformal field theory (CFT)}
\label{CFT}
Conformal field theory is quantum field theory with an extra symmetry that includes translations, Lorentz rotations, scale transformations, and special conformal transformations. We will first discuss CFT in higher dimensions [sec \ref{CFTdg3}] and then CFT in two dimensions in sec. \ref{CFTd2}. Lecture on CFT in higher dimensions is based on \cite{Qualls:2015qjb}.

\subsection{Conformal field theory in $d\geq 3$ dimensions} \label{CFTdg3}

Let us start with the $d$-dimensional space of signature $(p,q)$ as $\mathbb{R}^{p,q}$ (where $p+q=d$) equipped with flat metric $g_{\mu\nu}=\eta_{\mu\nu}=\text{diag}(-1,\dots,+1,\dots)$ and the line element
\bea
 ds^2 = g_{\mu\nu} dx^\mu dx^\nu. 
\eea 
 A differentiable map $\phi$ is defined as a conformal transformation if the metric transforms as 
 \bea \label{phi}
 \phi:g_{\mu\nu}(x)\rightarrow  g'_{\mu\nu}(x')=\Lambda(x)g_{\mu\nu}(x),
\eea 
  where $\Lambda(x)$ is known as scale factor. Metric transform as follows under the coordinate transformation $x\rightarrow x'$:
\bea  \label{gct}
  g_{\rho\sigma}\rightarrow g'_{\rho\sigma}(x') = \frac{\partial x'^{\mu}}{\partial x^{\rho}} \frac{\partial x'^{\nu}}{\partial x^{\sigma}}  g_{\mu\nu}(x).
  \eea
Therefore conformal transformations of the flat metric ($g_{\mu\nu}=\eta_{\mu\nu}$) using (\ref{phi}) and (\ref{gct}) is given by
\begin{equation}
\eta_{\rho\sigma}  \frac{\partial x'^{\rho}}{\partial x^{\mu}}\frac{\partial x'^{\sigma}}{\partial x^{\nu}} =\Lambda(x) \eta_{\mu\nu} . \label{eq:eq2p1}
\end{equation}
When $\Lambda(x)=1$, we obtain the Poincaré group, which consists of translations and Lorentz rotations. Further, when $\Lambda(x)$ is constant, we obtain global scale transformations. 

To study the conformal transformations, we start with the infinitesimal coordinate transformations up to first order in the parameter $\epsilon(x)\ll 1$ as follows
\begin{equation}
x'^\mu = x^\mu + \epsilon^\mu(x). \label{eq:eq2p2}
\end{equation}
Under infinitesimal coordinate transformation (\ref{eq:eq2p2}), left hand side of (\ref{eq:eq2p1}) is simplified as
\begin{eqnarray}
\eta_{\rho\sigma} \frac{\partial x'^{\rho}}{\partial x^{\mu}}\frac{\partial x'^{\sigma}}{\partial x^{\nu}}
&=& \eta_{\rho\sigma} \left(\delta^{\rho}_{\mu}+\frac{\partial \epsilon^\rho}{\partial x^{\mu}} + \mathcal{O}(\epsilon^2)  \right)   \left( \delta^{\sigma}_{\nu}+\frac{\partial \epsilon^\sigma}{\partial x^{\nu}} + \mathcal{O}(\epsilon^2) \right) \nonumber \\
&=& \eta_{\mu\nu} + \left( \frac{\partial \epsilon_{\mu}}{\partial x^\nu} + \frac{\partial \epsilon_{\nu}}{\partial x^\mu} \right) . \label{eq:eq2p3} 
\end{eqnarray}
Therefore we see that an infinitesimal transformation (\ref{eq:eq2p2}) will be a conformal transformation if in (\ref{eq:eq2p3})
\begin{equation}
\partial_\mu \epsilon_\nu + \partial_\nu \epsilon_\mu = f(x)\eta_{\mu\nu},
\label{eq:eq2p4}
\end{equation}
with $f(x)$ being some function and $\partial_\mu \equiv \frac{\partial}{\partial x^\mu}$. After taking the trace of both sides of (\ref{eq:eq2p4}) with $\eta^{\mu\nu}$, we get $f(x)=\frac2d \partial^\mu \epsilon_\mu$. Substituting obtained $f(x)$ back into (\ref{eq:eq2p4}), results in the following equation
\begin{equation}
\partial_\mu \epsilon_\nu + \partial_\nu \epsilon_\mu = \frac{2}{d}(\partial_\rho \epsilon^\rho) \eta_{\mu\nu}.
\label{eq:eq2p5}
\end{equation}
Using (\ref{eq:eq2p3}) and (\ref{eq:eq2p5}), we see that the conformal transformation for an infinitesimal coordinate transformation is given by\footnote{Due to Einstein summation convention: $\partial_\rho \epsilon^\rho=\partial_\mu \epsilon^\mu$.}
\begin{eqnarray}
\eta_{\rho\sigma} \frac{\partial x'^{\rho}}{\partial x^{\mu}}\frac{\partial x'^{\sigma}}{\partial x^{\nu}}
&=&  \left( 1  + \frac{2}{d}(\partial_\mu \epsilon^\mu)\right) \eta_{\mu\nu}. \label{eq:eq2p3-i} 
\end{eqnarray}
On comparing (\ref{eq:eq2p1}) and (\ref{eq:eq2p3-i}), we obtain the scale factor associated with the infinitesimal coordinate transformation as given below
$$\Lambda(x) = 1 + \frac2d (\partial_\mu \epsilon^\mu) .$$

We will now derive two expressions, using which we will see the structure of the conformal transformation later. Acting with $\partial^\nu$ on (\ref{eq:eq2p5})  gives the following equation
\begin{equation}
\partial_\mu (\partial \cdot \epsilon) + \Box \epsilon_\mu = \frac2d \partial_\mu (\partial \cdot \epsilon),
\label{eq:eq2p6}
\end{equation}
where $\partial \cdot \epsilon \equiv \partial_\mu \epsilon^\mu$ and $\Box\equiv\partial_\mu \partial^\mu$. Acting on (\ref{eq:eq2p6}) with $\partial_\nu$ gives the relation as below
\begin{equation}
\partial_\mu \partial_\nu (\partial \cdot \epsilon) + \Box \partial_\nu \epsilon_\mu = \frac2d \partial_\mu \partial_\nu (\partial \cdot \epsilon).
\label{eq:eq2p7}
\end{equation}
By the exchange of $\mu \leftrightarrow \nu$ in (\ref{eq:eq2p7}), we get
\begin{equation}
\partial_\nu \partial_\mu (\partial \cdot \epsilon) + \Box \partial_\mu \epsilon_\nu = \frac2d \partial_\nu \partial_\mu (\partial \cdot \epsilon).
\label{eq:eq2p7-i}
\end{equation}
Adding (\ref{eq:eq2p7}) and (\ref{eq:eq2p7-i}) and and using (\ref{eq:eq2p5}), we obtained the followng equation
\begin{equation}
\left( \eta_{\mu\nu} \Box + (d-2)\partial_\mu \partial_\nu \right) (\partial \cdot \epsilon) = 0.
\label{eq:eq2p8}
\end{equation}
Contracting equation (\ref{eq:eq2p8}) with $\eta^{\mu\nu}$ results in
\begin{equation}
(d-1) \Box (\partial \cdot \epsilon)=0.
\label{eq:eq2p9}
\end{equation}
Notice that as long as $d\geq 3$, (\ref{eq:eq2p8}) reduces to (\ref{eq:eq2p9}). For $d=2$, we will not get (\ref{eq:eq2p9}) from (\ref{eq:eq2p8}) because second term of (\ref{eq:eq2p8}) will be zero in two dimensions. We will focus on CFT in two dimensions in sec. \ref{CFTd2}.

Equation (\ref{eq:eq2p9}) suggests to us that $\epsilon(x)$ can be at most quadratic in $x$, which implies that $\partial \cdot \epsilon$ can be at most linear in $x$. This can be seen as follows
\bea
& & \Box (\partial \cdot \epsilon) \sim \partial_\mu \partial^\mu (x^\mu)= \partial_\mu (1) =0,
\eea
where we used $\Box=\partial_\mu \partial^\mu$ and $\partial^\mu (x^\mu)=\delta^\mu_\mu=1$. We will give the explicit form of $\epsilon(x)$ a little bit later. Before that, let us derive another expression, which will be useful for us.

 Act with $\partial_\rho$ on (\ref{eq:eq2p5}) and permute the indices to obtain the following set of equations
\begin{eqnarray}
\partial_{\rho} \partial_{\mu} \epsilon_\nu + \partial_{\rho} \partial_{\nu} \epsilon_\mu &=& \frac2d \eta_{\mu\nu}\partial_\rho (\partial \cdot \epsilon), \label{eq:eq2p1p1} \\
\partial_{\nu} \partial_{\rho} \epsilon_\mu + \partial_{\mu} \partial_{\rho} \epsilon_\nu &=& \frac2d \eta_{\rho\mu}\partial_\nu (\partial \cdot \epsilon), \label{eq:eq2p1p2}\\
\partial_{\mu} \partial_{\nu} \epsilon_\rho + \partial_{\nu} \partial_{\mu} \epsilon_\rho &=& \frac2d \eta_{\nu\rho}\partial_\mu (\partial \cdot \epsilon). \label{eq:eq2p1p3}
\end{eqnarray}
Adding (\ref{eq:eq2p1p2}) and (\ref{eq:eq2p1p3}) and then subtracting (\ref{eq:eq2p1p1}) from the summation of these two, we obtain the following expression 
\begin{equation}
2 \partial_\mu \partial_\nu \epsilon_\rho = \frac2d \left(  -\eta_{\mu\nu}\partial_{\rho}+\eta_{\rho\mu}\partial_\nu + \eta_{\nu\rho}\partial_\mu \right) (\partial \cdot \epsilon).
\label{eq:eq2p10}
\end{equation}

\subsubsection{Infinitesimal conformal transformations}
Let us come back to what we discussed earlier that $\epsilon_\mu$ can be at most quadratic in $x^\nu$, and the explicit form is written in terms of the coefficients $a_\mu,b_{\mu\nu},c_{\mu\nu\rho}\ll 1$ as below
\begin{equation}
\epsilon_\mu = a_\mu + b_{\mu\nu}x^\nu + c_{\mu\nu\rho}x^\nu x^\rho,
\label{eq:eq2p11}
\end{equation}
 $c_{\mu\nu\rho}=c_{\mu\rho\nu}$ because $x^\nu x^\rho=x^\rho x^\nu$. If we take $c_{\mu\nu\rho}$ as antisymmetric in its last two indices, then the third term in (\ref{eq:eq2p11}) will be zero, as the product of an antisymmetric and a symmetric tensor is zero. We will individually study each term in (\ref{eq:eq2p11}). Let us rewrite the infinitesimal coordinate transformation for the completeness for further discussion
 \bea \label{ict}
& & x'_\mu = x_\mu + \epsilon_\mu \, .
\eea
  %Because the constraints derived here for conformal invariance must be independent of the position $x^\mu$ (this should be a conformal transformation regardless of the value of $x^\mu$), the terms in equation ($\ref{eq:eq2p11}$) can be studied individually.
\begin{itemize}
\item {\bf Constant term $a_\mu$}: This corresponds to
\bea \label{ict-a}
& & x'_\mu = x_\mu + a_\mu
\eea
 This is a well-known infinitesimal translation whose generator is the momentum operator $P_\mu = -i\partial_\mu$.\\
 {\bf Excercise}: Derive the generator for infinitesimal translation.

\item {\bf Linear in $x$}: The infinitesimal coordinate transformation for the term linear in $x$ can be written as
\bea \label{ict-a-x}
& & x'_\mu = x_\mu + b_{\mu \nu} x^\nu
\eea
Substituting (\ref{ict-a-x}) in (\ref{eq:eq2p5}) gives
\begin{equation}
b_{\mu\nu}+b_{\nu\mu}=\frac2d \left( \eta^{\rho\sigma} b_{\rho\sigma} \right) \eta_{\mu \nu}.
\label{eq:eq2p12}
\end{equation}
The above equation constrains the symmetric part of $b$ to be proportional to the metric. Therefore $b_{\mu\nu}$ can be written as
\begin{equation}
b_{\mu\nu}=\alpha \eta_{\mu\nu}+m_{\mu\nu},
\label{eq:eq2p13}
\end{equation} 
where $m_{\mu\nu}=-m_{\nu\mu}$ that corresponds to infinitesimal Lorentz rotations $x'^{\mu}=(\delta^\mu_\nu+m^\mu_\nu)x^\nu$\footnote{This can be seen as: \bea \label{ict-a-x-as}
& & x'^\mu = x^\mu +  m^{\mu}_{\nu} x^\nu =(\delta^\mu_\nu+m^\mu_\nu)x^\nu.
\eea} with generators as the angular momentum operator: $L_{\mu\nu} = i(x_\mu\partial_\nu-x_\nu\partial_\mu)$. $\alpha$ is a parameter which can be obtained in terms of (\ref{eq:eq2p12}). The symmetric part of (\ref{eq:eq2p13}) simplifies equation (\ref{ict-a-x}) as  
\bea \label{ict-a-x-s}
& & x'_\mu = x_\mu + \alpha \eta_{\mu \nu} x^\nu =(1+\alpha)x^\mu.
\eea
The above transformation is an infinitesimal scale transformation with generators being $D=-ix^\mu \partial_\mu$.\\
{{\bf Excercise}: Derive the generators of infinitesimal Lorentz rotations and infinitesimal scale transformations.}\\
~\\
Let us summarize what we have obtained so far up to the term linear in $x$ in (\ref{eq:eq2p11}).
\bea \label{P+S}
& & x'_\mu = x_\mu + a_\mu \, ,\nn \\
& & x'^\mu  =(\delta^\mu_\nu+m^\mu_\nu)x^\nu \, , \nn \\
& & x'_\mu  =(1+\alpha)x^\mu \, .
\eea
From (\ref{P+S}), we see that the first two transformations correspond to the Poincaré group, whereas the third transformation corresponds to infinitesimal scale transformations. Now we will look at the term quadratic in $x$ in (\ref{eq:eq2p11}).

\item {\bf Quadratic in $x$}: Substituting $\epsilon_\mu$ from (\ref{eq:eq2p11}) in (\ref{eq:eq2p10}), we obtained the parameter $c_{\mu\nu\rho}$ in the following form
\begin{equation} \label{Cp}
c_{\mu\nu\rho}=\eta_{\mu\rho} b_\nu + \eta_{\mu\nu} b_{\rho} - \eta_{\nu\rho} b_{\mu}, \;\;\;\;\; b_\mu = \frac{1}{d} c^\nu_{\nu\mu}.
\end{equation}
Let us focus on 
\bea \label{SCT-d}
& & x'^\mu = x^\mu + c^{\mu\nu\rho} x_\nu x_\rho.
\eea
Substituting $c_{\mu\nu\rho}$ from (\ref{Cp}) in (\ref{SCT-d}), we obtain
\begin{equation} \label{SCTD}
x'^\mu = x^\mu + 2(x\cdot b)x^\mu - (x^2)b^\mu,
\end{equation}
where $x\cdot b=x^\mu b_\mu$ and $x^2=\eta^{\nu \rho}x_\nu x_\rho$. Transformations of the type (\ref{SCTD}) are known as {\it infinitesimal special conformal transformations} with generator as $K_\mu = -i (2x_\mu x^\nu \partial_\nu - (x^2) \partial_\mu)$. \\
{\bf Excercise}: Derive the generator of infinitesimal special conformal transformations. 

\end{itemize}

The finite conformal transformations corresponding to the infinitesimal conformal transformations are easy to imagine. For example, momentum, angular momentum, and dilatation generate translations, Lorentz rotations, and scale transformations, respectively. Now we will discuss how to imagine a special conformal transformation. Before that, we will provide the result for a finite special conformal transformation, which can be shown as an exercise. The finite special conformal transformation and the associated scale factor with it is given by
\begin{eqnarray}\label{eq:eq3p20}
& & x'^\mu = \frac{x^\mu-(x\cdot x) b^\mu}{1-2(b\cdot x)+(b\cdot b)(x \cdot x)}, \nonumber \\
& &  \Lambda(x) = \left( 1-2(b\cdot x) + (b\cdot b)(x \cdot x)  \right)^2.
\end{eqnarray}
{\bf Excercise}: Prove the above.\\
{\bf SCT justification}: First let us define \emph{inversions} as follows 
$$
x^\mu \rightarrow \frac{x^\mu}{x^2}.
$$
Inversion is a discrete transformation. The finite special conformal transformations using inversions can be written as 
\begin{equation}
\frac{x'^\mu}{x' \cdot x'} = \frac{x^\mu}{x \cdot x} - b^\mu.
\label{eq:eq2p71}
\end{equation}
{\bf Excercise}: Use $x'^\mu$ from (\ref{eq:eq3p20}) and prove the above.\\
Therefore, from (\ref{eq:eq2p71}), we can imagine special conformal transformations as an inversion of $x$ followed by a translation in $b$ and then another inversion.

\noindent In summary, we have the following generators for the infinitesimal conformal transformations
\begin{align}
&P_\mu = -i \partial_\mu \, , \nonumber \\
&L_{\mu\nu} = i(x_\mu\partial_\nu - x_\nu\partial_\mu) \, , \\
&D = -i x^\mu\partial_\mu \, , \nonumber \\
&K_\mu = -i(2x_\mu x^\nu \partial_\nu - x^2 \partial_\mu) \, . \nonumber
 \end{align} 
The algebra associated with infinitesimal conformal transformations is given below 
 \begin{align}
[D, P_{\mu}] &= \,i P_{\mu} \, , \nonumber \\
[D, K_{\mu}] &= -i K_{\mu} \, ,\nonumber \\
[K_{\mu}, P_\nu] &=  2 i ( \eta_{\mu\nu} D - L_{\mu\nu} ) \, , \label{eq:eq228}\\
[K_{\rho}, L_{\mu\nu}] &= i(\eta_{\rho\mu} K_{\nu}-\eta_{\rho\nu} K_{\mu}) \, , \nonumber \\
[P_{\rho}, L_{\mu\nu}] &= i(\eta_{\rho\mu}P_{\nu} - \eta_{\rho\nu}P_{\mu}) \, ,  \nonumber \\
[L_{\mu\nu}, L_{\rho\sigma}] &= i(\eta_{\nu\rho} L_{\mu\sigma} + \eta_{\mu\sigma}L_{\nu\rho} - \eta_{\mu\rho} L_{\nu\sigma} - \eta_{\nu\sigma} L_{\mu\rho}) \, . \nonumber
\end{align}
From (\ref{eq:eq228}), one can clearly see that one recovers Poincaré algebra if we ignore the commutators of $D$ and $K_\mu$.

\subsubsection{Conformal group and scaling dimension}
{\bf Conformal group}: Now we will discuss the conformal group for $d\geq 3$. The Lie algebra associated with the conformal group is the conformal algebra. The number of generators associated with the conformal algebra are counted as follows
\begin{eqnarray}
1 \text{  dilatation} \;\;&+&\;\; d \text{  translations} +\;\; d \text{  special conformal transformations} \nonumber \\
 &+& \frac{d(d-1)}{2} \text{  rotations} = \frac{(d+2)(d+1)}{2} \text{  generators}. \nonumber
\end{eqnarray}
These are the same as the number of generators of the algebra associated with the $SO(d+2)$ group. Let us define alternate generators as 
\begin{eqnarray} \label{Jgen}
J_{\mu,\nu} &\equiv& L_{\mu\nu} \, , \nonumber \\
J_{-1,\mu}&\equiv& \frac12 (P_\mu-K_\mu) \, ,  \nonumber \\
J_{0,\mu}&\equiv& \frac12 (P_\mu+K_\mu) \, , \\ 
J_{-1,0}&\equiv& D \, .\nonumber
\end{eqnarray}
These generators satisfy the following algebra
\begin{equation}
\left[ J_{mn},J_{pq}   \right] = i \left( \eta_{mq}J_{np} + \eta_{np}J_{mq} - \eta_{mp}J_{nq} - \eta_{nq}J_{mp} \right). \label{eq:129}
\end{equation}
We can embed the $d$ dimensional Euclidean space $\mathbb{R}^{d,0}$ in $(d+2)$ dimensional Minkwoski spacetime with metric signature diag$(-1,1,\dots,1)$. The commutation relations (\ref{eq:129}) are the Lie algebra of $SO(d+1,1)$ for Euclidean space. Similarly, $d$ dimensional Lorentzian spacetime $\mathbb{R}^{d-1,1}$ is embedded in $(d+2)$ dimensional Minkwoski spacetime with metric signature diag$(-1,-1,1,\dots,1)$ and the algebra (\ref{eq:129}) corresponds to the Lie algebra of $SO(d,2)$ group\footnote{Conformal algebra for $d=p+q$ is $so(p+1,q+1)$.}. \\
 {\bf Excercise}: Prove (\ref{eq:129}).

{\it We conclude that the conformal group for Euclidean and Lorentzian CFTs in $d$ dimensions (with $d\geq3$) is $SO(d+1,1)$ and $SO(d,2)$ respectively. Later, when we discuss AdS$_{d+1}$ spacetime in sec. \ref{AdSmd}, we will see that there is a matching of the symmetry group of AdS with the conformal group.}

{\bf Scaling dimension}: If a field $\Phi$ transforms according to the equation given below (with scale transformation $x \rightarrow \lambda x$) 
\begin{equation} \label{scd}
\Phi(\lambda x) = \lambda^{-\Delta} \Phi(x).
\end{equation}
Then $\Delta$ is known as the scaling dimension of the field $\Phi$. We can determine the scaling dimension by demanding the invariance of action under (\ref{scd}) and scale transformation $x \rightarrow \lambda x$. Let us consider the action corresponding to a free massless scalar field $\Phi(x)$ in flat space as
\begin{equation} \label{Action}
S = \int d^d x \;\;\partial_\mu \Phi(x) \partial^\mu \Phi(x).
\end{equation}
We can show that under (\ref{scd}) and  $x \rightarrow \lambda x$, the action (\ref{Action}) is invariant if the scaling dimension of $\Phi$ is
\begin{equation}
\Delta = \frac{d}{2} - 1.
\end{equation}

\subsubsection{Two point function of CFT$_d$}
\label{twoptcft}
Here we will derive the two-point function of a scalar field in conformal field theory in $d$ dimensions based on the symmetries of CFT. Later in sec. \ref{twoptAdS}, we will show that this can be derived from AdS bulk using $AdS/CFT$ correspondence.

As we discussed earlier, under dilatations: $x \rightarrow \lambda \, x$, scalar field transform as
\bea
& & \phi(\lambda  x)= \lambda^{-\Delta}\, \phi(x).
\eea
Therefore, two point function under dilatations behaves as
\begin{equation}
\langle \phi_1(x_1) \phi_2(x_2) \rangle = \lambda^{\Delta_1+\Delta_2}\langle \phi_1(\lambda x_1) \phi_2(\lambda x_2) \rangle.
\end{equation}
Therefore a function $f(x)$ will behave as
\begin{equation}
f(x)=\lambda^{\Delta_1+\Delta_2} f(\lambda x).
\end{equation}
One can show that Poincar\'{e} invariance (consists of translations and Lorentz rotations) requires
\begin{equation}
\langle \phi_1(x_1) \phi_2(x_2) \rangle = f(|x_1-x_2|).
\end{equation}
As $x_1-x_2$ is invariant under $x_{1,2} \rightarrow x_{1,2}+ \epsilon$ where $\epsilon$ is some constant parameter, and we know that rotation will be preserved because of $|x_1-x_2|$. 
From the above discussions with the invariance under dilatations and Poincaré invariance, we can argue that the form of two point function will be as written below 
\begin{equation} \label{a2}
\langle \phi_1(x_1) \phi_2(x_2) \rangle = \frac{d_{12}}{|x_1-x_2|^{\Delta_1+\Delta_2}},
\end{equation}
with $d_{12}$ being some normalization constant which depends on the fields $\phi_1,\phi_2$. This is the only form with the appropriate transformation properties.

Quasi-primary fields transform as
\bea
& & \phi(x) \rightarrow \phi(x')=\left\lvert \frac{\partial x'}{\partial x} \right\rvert^{- \Delta /d } \, \phi(x).
\eea
 For a special conformal transformation,
\begin{equation}
\left\lvert \frac{\partial x'}{\partial x} \right\rvert =\frac{1}{\gamma^d},
\end{equation}
where $\gamma=(1-2b\cdot x + b^2 x^2)$. Therefore
\bea
& &  \phi(x')=\gamma^{\Delta} \, \phi(x) \, ,
\eea
and hence
\begin{equation} \label{a1}
\langle \phi_1(x_1) \phi_2(x_2) \rangle =\frac{1}{\gamma_1^{\Delta_1}\gamma_2^{\Delta_2}}\langle \phi_1(x_1') \phi_2(x_2') \rangle.
\end{equation}
Under the special conformal transformation (\ref{eq:eq3p20}), one can show that
\begin{equation} \label{dsct}
|x_i'-x_j'| = \frac{|x_i-x_j|}{(1-2b\cdot x_i + b^2 x_i^2)^{1/2}(1-2b\cdot x_j + b^2 x_j^2)^{1/2}}.
\end{equation}
Hence from (\ref{a1}), (\ref{a2}) and (\ref{dsct}), we have
\begin{eqnarray}
& & \hskip -1.1in \langle \phi_1(x_1) \phi_2(x_2) \rangle =\frac{1}{\gamma_1^{\Delta_1}\gamma_2^{\Delta_2}}\langle \phi_1(x_1') \phi_2(x_2') \rangle \nn \\
& & =\frac{1}{\gamma_1^{\Delta_1}\gamma_2^{\Delta_2}}\frac{d_{12}}{|x_1'-x_2'|^{\Delta_1+\Delta_2}} \nn \\
& & = \frac{d_{12}}{\gamma_1^{\Delta_1}\gamma_2^{\Delta_2}} \frac{(\gamma_1 \gamma_2)^{(\Delta_1+\Delta_2)/2}}{|x_1-x_2|^{\Delta_1+\Delta_2}},
\end{eqnarray}
with $\gamma_i \equiv (1-2b\cdot x_i + b^2 x_i^2)$. The above constraint is true only if $\Delta_1=\Delta_2$. To summarise, the two point function invariant under dilatations, Poincaré, and SCT has the following form
 
\begin{eqnarray} \label{twoptcftf}
& & \hskip -1.08in \langle \phi_1(x_1) \phi_2(x_2) \rangle =
\frac{d_{12}}{|x_1-x_2|^{2\Delta_1}} \ \ \ \  \text{if } \Delta_1 = \Delta_2 \nn \\
& & =0 \ \ \ \ \ \ \ \ \ \ \ \ \ \ \ \ \ \ \ \text{if } \Delta_1 \neq \Delta_2.
\end{eqnarray}

\subsection{Conformal field theory in two dimensions}
\label{CFTd2}
Here, we will discuss the conformal field theory in two dimensions. We are following \cite{Qualls:2015qjb} \cite{Nawata:2022ywk} for discussion on CFT in two dimensions.

Let's start with some basics of complex analysis which is required for CFT in two dimensions. Let $f(z)$ be a complex function defined as
\bea
& & f(z)=u(x,y)+ i v(x,y)
\eea
where $u(x,y)$ and $v(x,y)$ are real valued functions. Then $f(z)$ is differentiable at a point $z_0=x_0+i y_0$ in the complex plane iff partial derivatives of $u$ and $v$ exist and are continuous in the neighbourhood of $z_0$, and they satisfy the Cauchy-Riemann (C-R) equations as given below
\bea \label{CR}
& & \frac{\partial u}{\partial x}=\frac{\partial v}{\partial y} \, , \ \ \ \ \ \frac{\partial u}{\partial y}=-\frac{\partial v}{\partial x}.
\eea
The Cauchy-Riemann equations will appear in two-dimensional CFT, which we will see below. To see this, consider CFT in two-dimensional Euclidean space with metric
\bea
& & ds^2=\eta_{\mu \nu} dx^\mu dx^\nu ; \ \ \ \ \ \eta_{\mu \nu}=\begin{pmatrix}
1 & 0  \\
0 & 1
\end{pmatrix}.
\eea
Let us rewrite equation (\ref{eq:eq2p5}) here for the convenience, which is useful for the current discussion
\bea \label{ME}
& & \partial_\mu \epsilon_\nu +\partial_\nu \epsilon_\mu = \frac{2}{d}\left(\partial \cdot \epsilon\right) \eta_{\mu \nu} \, .
\eea
Simplifying the above equation component-wise results in the following equations
\bea
& & [\mu=\nu=1] \ \ \ \ \ \partial_1 \epsilon_1= \frac{2}{d}\left(\partial \cdot \epsilon\right) \eta_{11}, \nn \\
& & [\mu=\nu=2] \ \ \ \ \ \partial_2 \epsilon_2= \frac{2}{d}\left(\partial \cdot \epsilon\right) \eta_{22}, \nn \\
& & [\mu=1, \nu=2] \ \ \ \ \ \partial_1 \epsilon_2+ \partial_2 \epsilon_1= 0
\eea
which implies
\bea \label{CR-ep}
& & \partial_1 \epsilon_1=\partial_2 \epsilon_2 \, ; \ \ \ \ \ \ 
 \partial_1 \epsilon_2=- \partial_2 \epsilon_1.
\eea 
The above equations are the analog of C-R equations (\ref{CR}) in the complex plane. We introduce the complex coordinates $z$ and $\overline{z}$, parameters $\epsilon$ and $\overline{\epsilon}$ as
\bea
& & z=x^1+ i x^2 ,  \ \ \ \  \overline{z}=x^1- i x^2, \nn \\
& & \epsilon=\epsilon_1 + i \epsilon_2, \ \ \ \ \ \ \overline{\epsilon}=\epsilon_1 - i \epsilon_2 \, .
\eea
Equation (\ref{CR-ep}) implies that $\epsilon(z)$ is a holomorphic function in some open set. This leads to the fact that $\epsilon(z)$ is a complex differential at every point in that open set. Therefore, any infinitesimal holomorphic transformation $z'=z+\epsilon(z)$ gives rise to a two-dimensional conformal transformation. We can rewrite equation (\ref{ME}) in terms of the complex coordinate as
\bea
& &  \Box \epsilon_\mu =0
\eea
where $\Box = \partial^\nu \partial_\nu=\partial \overline{\partial}$ with the notation $\partial = \partial_z, \bar{\partial} = \partial_{\bar{z}}$. In complex coordinate, the metric of two dimensional Euclidean space becomes
\bea
& & ds^2=dx_1^2 + d x_2^2=\frac{1}{2} dz d\overline{z}; \ \ \ \ \ \ g_{\alpha \beta}=\begin{pmatrix}
0 & \frac{1}{2}  \\
\frac{1}{2} & 0
\end{pmatrix}.
\eea

Using these coordinates, the holomorphic Cauchy-Riemann equations become
\begin{equation}
\bar{\partial} f(z,\bar{z}) = 0.
\end{equation}
Therefore, any holomorphic function in the complex plane is conformal. The solution to this equation is any holomorphic mapping
$$
z\rightarrow f(z).
$$
Hence, the conformal group in two dimensions consists of a set of all analytic maps, which is infinite-dimensional, associated with the coefficients of a Laurent series (we have discussed Laurent series in sec. \ref{WVA}) required for the analyticity of a function in the neighborhood. An infinitesimal conformal transformation in two dimensions is of the form $f(z)=z+\epsilon(z)$ and $f(\overline{z})=\overline{z}+\epsilon(\overline{z})$.
Metric transforms as follows
$$
ds^2 = dz d\bar{z} \rightarrow \frac{\partial {f}}{\partial {z}} \frac{\partial \bar{f}}{\partial \bar{z}} dz d\bar{z}. 
$$
Which leads to the conformal factor as $\Lambda = \left| \frac{\partial f}{\partial z} \right|^2$.

\subsubsection{Global conformal transformations}
We can generate infinitely many infinitesimal conformal transformations. However, for these transformations to form a group, they must be invertible and map the entire complex plane, including the point at infinity, onto itself. The transformations satisfying these requirements are called global conformal transformations, and together they constitute the special conformal group.

Explicitly, the global conformal transformations take the form
\begin{equation}
f(z)=\frac{az+b}{cz+d}.
\end{equation}
For this map to be invertible, the determinant $ad-bc$ must be nonzero. By convention, one fixes the overall scaling freedom of the coefficients by requiring $ad-bc=1$ since multiplying $a,b,c,d$ by the same constant does not alter the transformation. Therefore, the special conformal group in two dimensions is given by the projective transformations. For each transformation, we can describe it using the matrix\footnote{We will discuss in detail at the end of sec. \ref{WVA}.}
\begin{equation}
\left( \begin{array}{cc}
a & b \\
c & d
\end{array} \right),\;\;a,b,c,d\in\mathbb{C}.
\end{equation}
This implies that the global conformal group for two-dimensional CFTs is isomorphic to the group $SL(2,\mathbb{C})$, which itself is isomorphic to $SO(3,1)\sim SO(2,2)$ (Lorentz group in four dimensions).

\subsubsection{The Witt and Virasoro algebras}
\label{WVA}
Before starting this section, we would like to remind the readers definitions of Meromorphic function and Laurent series from complex analysis that will be useful for this section.\\
 {\bf Meromorphic function}: It is a generalization of the holomorphic function, which is allowed to have certain types of singularities called poles. The more elaborate definition is as follows. A function $f(z)$ is meromorphic in some open set $U \in \mathbb{C}$ if\\
(i) $f(z)$ is holomorphic in $U$ except at set of isolated points $\{ z_k \}$.\\
(ii) At each $z_k$, $f(z)$ has a pole of finite order.\\
$f(z)$ can be written locally near any pole $z_k$ as
\bea \label{LS}
& & f(z)=\frac{a_{-n}}{(z-z_k)^{n}}+ ......+\frac{a_{-1}}{(z-z_k)}+\sum_{m=0}^{\infty} a_m \, (z-z_k)^m \, ,
\eea
where $a_{-n} \neq 0$. The series (\ref{LS}) is known as the {\it Laurent series}.

In the begining of sec. \ref{CFTd2}, we discussed that two-dimensional conformal field theory admits an infinitesimal conformal transformation that is holomorphic in some open set. In general, let us assume that the infinitesimal conformal transformation $\epsilon(z)$ is a meromorphic function, and we can perform the Laurent series expansion around $z=0$ as follows
\begin{eqnarray}
{z}' = {f(z)} = {z}+{\epsilon}(z) = {z} + \sum_{n\in \mathbb{Z}} {\epsilon}_n \left( -{z}^{n+1}  \right), \nonumber \\ 
\bar{z}' = \bar{f}(\bar{z}) = \bar{z}+\bar{\epsilon}(\bar{z}) = \bar{z} + \sum_{n\in \mathbb{Z}} \bar{\epsilon}_n \left( -\bar{z}^{n+1}  \right),
\label{eq:eq3p3}
\end{eqnarray}
where parameters $\epsilon_n$ and $\bar{\epsilon}_n$ are infinitesimal and constant. Let us derive the generator for this infinitesimal transformation. Let $\phi(z,\bar{z})$ be a scalar and dimensionless field living on the plane. Under infinitesimal transformation, we demand that: $\phi(z,\bar{z}) \rightarrow \phi'(z',\bar{z}')=\phi(z,\bar{z})$ and
\bea
& & \hskip -0.2in \delta \phi=\phi'(z,\bar{z})-\phi(z,\bar{z}) \nn \\
& & =\phi'(z'-\epsilon(z),\bar{z}'-\bar{\epsilon}(z))-\phi(z,\bar{z}) \nn \\ 
& & =\phi'(z',\bar{z}')- \epsilon(z) \partial \phi'(z',\bar{z}')-\bar{\epsilon}(\bar{z}) \bar{\partial} \phi'(\bar{z}',\bar{z}')-\phi(z,\bar{z}) \nn \\
& & =- \epsilon(z) \partial \phi(z,\bar{z})-\bar{\epsilon}(\bar{z}) \bar{\partial}\phi(z,\bar{z}) \nn \\
& & =- \sum_{n \in \mathbb{Z}}\left(\epsilon_n \ell_n \phi(z,\bar{z})+\bar{\epsilon}_n \bar{\ell}_n \phi(z,\bar{z})\right),
\eea
where $l_n= - z^{n+1} \partial$ and $\bar{l}_n= - \bar{z}^{n+1} \bar{\partial}$ are the generators of infinitesimal transformation for two-dimensional conformal field theory. Thus, there are infinitely many infinitesimal conformal transformations in two dimensions. The conformal algebra associated with these generators is given by
\begin{align} \label{W-A}
[\ell_m, \ell_n] &=  (m-n)\ell_{m+n}, \nonumber \\
[\bar{\ell}_m, \bar{\ell}_n] &= (m-n)\bar{\ell}_{m+n} , \\
[\ell_m, \bar{\ell}_n] &= 0. \nonumber
\end{align}
{\bf Excercise}: Derive the above algebra.\\
~\\
We have two copies of Witt algebra in the first and second lines of (\ref{W-A}). There is a finite subalgebra in two dimensions generated by $\ell_{-1}$, $\ell_0$, and $\ell_1$. The aforementioned subalgebra is associated with the global conformal group.

The generators on $\mathbb{R}^2\simeq\mathbb{C}$ are not defined everywhere. Instead, we work with the Riemann sphere $S^2\sim\mathbb{C}\cup\{\infty\}$, which includes a point at infinity. $\ell_n$ generators are well defined (non-singular) at $z=0$ only for $n\geq-1$. To see the behaviour at $z \rightarrow \infty$, let us define $z=- \frac{1}{w}$. In the $w$ variabble, $\ell_n$ is given as: $\ell_n=\left(-\frac{1}{w} \right)^{n-1}\partial_w$ which is non-singular at $w=0 (z \rightarrow \infty)$ for $n\leq1$. This discussion suggests that global conformal transformations on the Riemann sphere are generated by $\ell_{-1}, \ell_0,$ and $\ell_1$.

{\bf How to see the momentum, rotation, etc, as generators from this finite subalgebra?} Let us write the expressions first
\bea \label{ls-2d}
& & l_{-1}=- \partial_z, \ \ \ \ \ {\bar{l}}_{-1}=- \partial_{\bar{z}} ,\nn \\
& & l_{0}=- z \,  \partial_z, \ \ \ \ \ {\bar{l}}_{0}=-\bar{z} \, \partial_{\bar{z}} ,\nn \\
& & l_{1}=- z^2 \, \partial_z, \ \ \ \ \ {\bar{l}}_{1}=- \bar{z}^2 \, \partial_{\bar{z}} ,\nn \\
\eea
From (\ref{ls-2d}), we can see clearly that $l_{-1}$ and ${\bar{l}}_{-1}$ generates tanslations as $z'=z+a$ and $\bar{z}'=\bar{z}+\bar{a}$ which are generators of momentum. Let us define: $z=r \,  e^{i \theta}$ then
\bea \label{ls-2d-i}
& & l_{0}=- \frac{1}{2} r \,  \partial_r + \frac{i}{2} \partial_\theta, \ \ \ \ \ {\bar{l}}_{0}=- \frac{1}{2} r \,  \partial_r -\frac{i}{2} \partial_\theta \, .
\eea 
Using (\ref{ls-2d-i}), we find that $\ell_0+{\bar{\ell}}_{0}=-r \,  \partial_r$ generates dilatations and $i\left(\ell_0-{\bar{\ell}}_{0}\right)=-  \partial_\theta$ generates rotations. The generators $\ell_1$ and $\bar{\ell}_1$ generate special conformal transformations. To see this, let us use $w$ variable: $z=-\frac{1}{w}$, in terms of $w$, we have $\ell_1=-\partial_w$ which generates translations as $w'=w-\epsilon_c$ and hence $-\frac{1}{z'}=-\frac{1}{z}-\epsilon_c$ implies $z'=\frac{z}{\epsilon_c \, z +1}$ which is the generator of special conformal transformations. We can argue similarly for $\bar{\ell}_1$. All of these operators generate transformations of the following form
\bea \label{SL}
& & z\rightarrow \frac{az+b}{cz+d}, \;\;\;\;\;\;a,b,c,d\in \mathbb{C}.
\eea
This is the transformation associated with the conformal group $SL(2,\mathbb{C})$. From (\ref{SL}), we see that under: $(a,b,c,d) \rightarrow (-a,-b,-c,-d)$, we recover the same result. Therefore the actual group is $SL(2,\mathbb{C})/\mathbb{Z}_2$ which is $PSL(2,\mathbb{C})$. Now let us write the explicit conformal transformations from (\ref{SL}):
\bea
& & {\rm Translation}: \ \ \ \ z \rightarrow z + b  \iff (a,b,c,d) \rightarrow (1,b,0,1) \, , \nn \\
& &  {\rm Dilatation}: \ \ \ \ z \rightarrow a \, z    \iff (a,b,c,d) \rightarrow (\sqrt{a},0,0,\frac{1}{\sqrt{a}}) \, , \nn \\
& &  {\rm Rotation}: \ \ \ \ z \rightarrow z\, e^{i \theta}  \iff (a,b,c,d) \rightarrow (e^{i \theta/2},0,0,e^{-i \theta/2}) \, , \nn \\
& &  {\rm Special \ conformal \ transformation}: \ \ \ \ z \rightarrow \frac{z}{c \, z + 1}  \iff (a,b,c,d) \rightarrow (1,0,0,1) \,. \nn \\
& & 
\eea
The central extension of the Witt algebra is the Virasoro algebra, as given below:
\begin{equation} \label{VL}
[L_m,L_n]= (m-n)L_{m+n} + \frac{c}{12}(m^3-m)\delta_{m+n,0},
\end{equation}
where $c$ is the central charge and there is similar algebra for $\bar{L}$ and $\bar{c}$. From the second term of (\ref{VL}) we see that a finite subalgebra of conformal transformations is unaffected by the central extension of the Witt algebra as the second term goes to zero for $m=(-1,0,1)$.

\section{Anti de Sitter (AdS) spacetime}
\label{AdSmd}
This section is based on \cite{Kaplan,Nastase}. Here we will discuss AdS spacetime in different coordinate systems with Lorentzian (sec. \ref{LAdS}) and Euclidean (sec. \ref{EAdS}) versions. We will also discuss the symmetry group of AdS spacetime. We end the AdS spacetime lecture with a discussion on the Poincaré patch of AdS spacetime in sec. \ref{PAdS-sec}.

\subsection{Lorentzian AdS spacetime} \label{LAdS}
\begin{figure}[h]
\begin{center}
\includegraphics[width=0.5\textwidth]{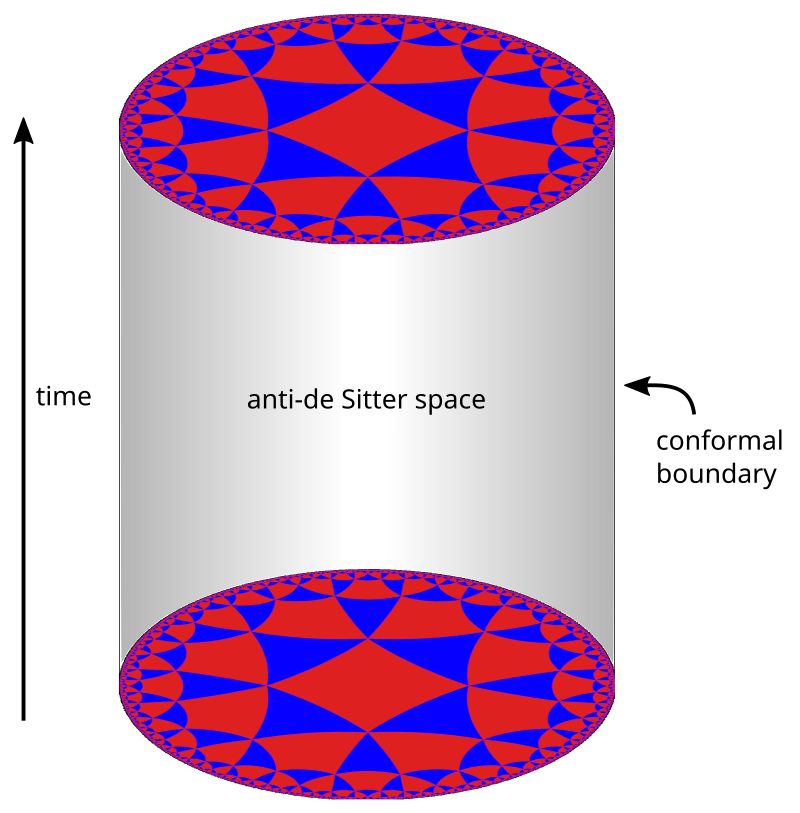}
\end{center}
\caption{Global AdS spacetime. $\rho$ is along the horizontal direction in this picture and $S^{d-1}$ is located at top and bottom. This picture is taken from \url{https://en.wikipedia.org/wiki/Anti-de_Sitter_space}.}
\label{AdS}
\end{figure}

AdS spacetime is a maximally symmetric spacetime that is a solution to the Einstein equations of motion with a negative cosmological constant.
The gravitational action is given by
\begin{equation}
 S=\frac{1}{16 \pi G_{d+1}} \int d^{d+1}x \sqrt{-g}\left(R- 2 \Lambda \right)\, .
\end{equation}
Equation of motion for the metric is obtained by varying the above equation with respect to the metric, which leads to
\begin{equation} \label{EE-Eq}
R_{\mu \nu}-\frac{1}{2}g_{\mu \nu}\left(R - 2 \Lambda\right)=0.
\end{equation}
In the global coordinate, metric (solution of (\ref{EE-Eq})) in the Lorentzian signature has the following form
\begin{equation} \label{M-gc}
ds^{2}=\frac{1}{\cos^{2}\left(\frac{\rho}{L} \right)}\left[-dt^{2}+d\rho^{2}+\sin^{2}\left(\frac{\rho}{L} \right)\, d\Omega_{d-1}^{2}\right]\ ,
\end{equation}
where $t \in (-\infty, \infty)$, $\rho \in [0,\frac{\pi}{2}L]$, $d\Omega_{d-1}^{2}$ is the metric of $S^{d-1}$ and $L$ is AdS length scale. At the conformal boundary, $\rho=\frac{\pi}{2}L$ , metric (\ref{M-gc}) is conformal to a solid cylinder with topology $\mathbb{R}\times S^{d-1}$. AdS can be viewed as the interior of a cylinder in global coordinates with center at $\rho=0$ and spatial infinity located at $\rho=\frac{\pi}{2}L$ as shown in Fig. \ref{AdS}.

The symmetry group of Lorentzian $AdS_{d+1}$ is $SO(2,d)$ which has $\frac{(d+1)(d+2)}{2}$ generators. This is same as Lorentzian CFT in $d$ dimensions (sec. \ref{CFTdg3})!! This can be seen by embedding $AdS_{d+1}$ into flat $d+2$ dimensional Minkowski spacetime ($M_{d+2}$) using the following embedding equation
 \begin{equation} \label{ES-AdS-L}
 X_A X^A=-X_0^2-X_1^2 +\sum_{i=1}^d X_i^2=-L^2.
  \end{equation}
Metric for $M_{d+2}$ is given as
\begin{equation} \label{M-L}
ds^2=-d X_0^2-d X_1^2 +\sum_{i=1}^d dX_i^2.
\end{equation} 
We can easily see that both equations (\ref{ES-AdS-L}) and (\ref{M-L}) are invariant under the group $SO(2,d)$ and number of generators associated with this group is $\frac{(d+1)(d+2)}{2}$. Generators are given as: $L^A_{B} = X^A \frac{\partial}{\partial X^B}-X^B \frac{\partial}{\partial X^A}$. Metric (\ref{M-gc}) can be obtained using (\ref{M-L}) with the use of following embedding coordinates
\bea \label{ESC}
& & X_0= L \frac{\cos t}{\cos \rho},\nn \\
& & X_{d+1}= L \frac{\sin t}{\cos \rho}, \nn \\
& & X_i = L \tan \rho \ \hat{\Omega}_i \, ,
\eea
where $\hat{\Omega}_i$ are the coordinates on $S^{d-1}$. Another representation of global $AdS_{d+1}$ is obtained by the following embedding coordinates (in $L=1$ units)
\bea \label{ESC-i}
& & X_0= L \cos t \cosh r,\nn \\
& & X_{d+1}= L \sin t \cosh r, \nn \\
& & X_i = L \sinh r \ \hat{\Omega}_i. 
\eea
Now using (\ref{M-L}), metric is obtained as
\bea \label{M-gc-i}
& & ds^2 = L^2 \left[ - \cosh^2 r \, dt^2 +dr^2 + \sinh^2 r \ d\Omega_{d-1}^{2} \right],
\eea
with $t \in [0, 2 \pi]$ and $r \geq 0$. Since $t$ is compact here, we use $\tanh r=\sin \rho$ to obtain (\ref{M-gc}) to see the conformal structure of AdS spacetime and noncompact time coordinate.\\

\textbf{Static coordinate system}: We can write the metric of AdS spacetime (\ref{M-gc-i}) in the static coordinate system (similar to de Sitter space which we will discuss later in sec. \ref{dSholography}) using the transformation $\tilde{r}=\sinh r$. To see this, let us focus on two dimensions with the global coordinate metric
\bea \label{M-gc-i-2D}
& & ds^2 = L^2 \left[ - \cosh^2 r dt^2 +dr^2 \right],
\eea
Now using $\tilde{r}=\sinh r$, (\ref{M-gc-i-2D}) becomes
\bea \label{M-AdS-sc}
& & ds^2= L^2 \left[-(\tilde{r}^2+1) dt^2 + \frac{d \tilde{r}^2}{(\tilde{r}^2+1)}\right].
\eea
The solution (\ref{M-AdS-sc}) looks like a black hole solution and a static patch of de Sitter space. The boundary of AdS spacetime in static coordinates is located at $\tilde{r} \rightarrow \infty $.

\subsection{Euclidean AdS background} \label{EAdS}
Euclidean $AdS_{d+1}$ can be embedded in $(d+2)$ dimensional Minkowski spacetime ($M_{1, d+1}$) with the following embedding equation 
\begin{equation}
-\left(X^{0}\right)^{2}+\left(X^{1}\right)^{2}+\dots+\left(X^{d+1}\right)^{2}=-L^{2}\ ,\qquad X^{0}>0\ .\label{eq:EAdS}
\end{equation}
 The metric associated withe $M_{1, d+1}$ is
\begin{equation} \label{M-L-E}
ds^2=-d X_0^2 +\sum_{i=1}^{d+1} dX_i^2.
\end{equation} 
We see that equations (\ref{eq:EAdS}) and (\ref{M-L-E}) are invariant under the group $SO(1,d+1)$, which is the same as the conformal group of Euclidean CFT in $d$ dimensions. {Generators of Euclidean conformal group is: $J_{AB}=-i\left(X_A \frac{\partial}{\partial X^B}-X_B \frac{\partial}{\partial X^A}\right)$.} Metric of Euclidean $AdS_{d+1}$ in global coordinate system is obtained by using the following embedding coordinates
\begin{eqnarray}
X^{0} & = & L \cosh\tau_E \cosh\rho \, ,\nonumber \\
X^{i} & = & L \,\Omega^{i}\sinh\rho \, , \label{eq:AdSEmbglobalcoord}\\
X^{d+1} & = & L \sinh\tau_E \cosh\rho \, .\nonumber 
\end{eqnarray}
On substiuting (\ref{eq:AdSEmbglobalcoord}) into (\ref{M-L-E}), one obtains the metric for global $AdS_{d+1}$ in the Eunclidean signature as written below
\be \label{M-EGAdS}
ds^{2}=L^{2}\left[\cosh^{2}\rho\, d\tau_E^{2}+d\rho^{2}+\sinh^{2}\rho\, d\Omega_{d-1}^{2}\right]\ ,
\ee
with $\tau_E \in [0, 2 \pi]$ and $\rho \geq 0$. To study the conformal structure of Euclidean global AdS, we use $\tanh \rho =\sin \theta$, which simplifies metric (\ref{M-EGAdS}) to the following form
\bea \label{M-EGAdS-i}
& & ds^2=\frac{L^2}{\cos^2 \theta}\left[d\tau_E^2 + d\theta^2 +\sin^2 \theta d\Omega_{d-1}^2 \right].
\eea
with $\theta \in [0, \pi/2]$. From the metric (\ref{M-EGAdS-i}), we can see that the conformal boundary of Euclidean $AdS_{d+1}$ is located at $\theta=\pi/2$ and constant Euclidean time slices are balls at fixed height on the cylinder.

\subsection{Poincaré patch of AdS spacetime}
\label{PAdS-sec}
To discuss the Poincaré patch of Lorentzian $AdS_{d+1}$, we follow specifically \cite{Bayona:2005nq}. There are also other references available in the literature. The Poincaré patch divides the global AdS into two parts, which we will discuss in this section. Before that, let us derive the metric of the Poincaré patch using the embedding space formalism. Embedding space coordinates for this purpose are given as
\bea \label{ESCPAdS}
& & X_0=\frac{1}{2 z}\left(z^2+L^2+ \vec{x}^2-t^2\right), \nn \\
& & X_i = \frac{L}{z} x^i,\nn \\
& & X_d=\frac{1}{2 z}\left(z^2-L^2+ \vec{x}^2-t^2\right), \nn \\
& & X_{d+1}= \frac{L \, t}{z}.
\eea
Using (\ref{ESCPAdS}), the metric for the Poincaré patch of AdS, similar to global AdS, is obtained as below
\bea
& & ds^2=\frac{L^2}{z^2}\left(- dt^2+dz^2+d\vec{x}^2\right).
\eea
\begin{figure}[h]
\begin{center}
\includegraphics[width=0.65\textwidth]{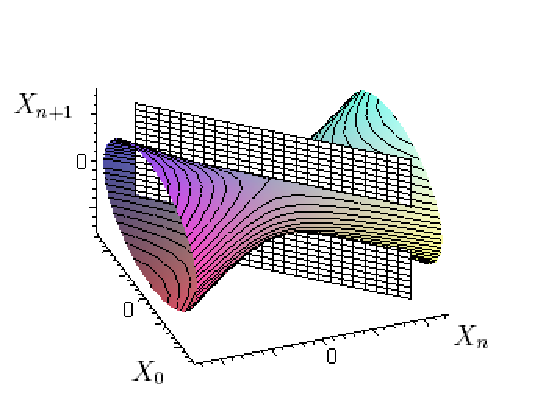}
\end{center}
\caption{Poincaré  patch of $AdS_{d+1}$ spacetime. This figure is taken from \cite{Bayona:2005nq}. To make it consistent with the current discussion, we need to replace $n$ with $d$ in this figure.}
\label{PAdS}
\end{figure}
Let us define the light cone coordinates $(u, v)$ and $z$ in terms of $u$ as
\bea
& & u=\frac{X_0-X_d}{L^2}, \ \ \ \ \ \ \ v=\frac{X_0+X_d}{L^2}, \ \ \ \ \ \ \ z=\frac{1}{u}=\frac{L^2}{X_0-X_d}.
\eea
Therefore, we have two different Poincaré charts
\begin{itemize}
\item $z>0$ which corresponds to $X_0 > X_d$.
\item $z<0$ which corresponds to $X_0 < X_d$.
\end{itemize}
The Poincaré patch of AdS is the region corresponding to one of these two charts. These two charts are divided by the hypersurface $X_0=X_d$, and this corresponds to $z \rightarrow \pm \infty$ limits. See Fig. \ref{PAdS} for pictorial representation.

\section{The AdS/CFT correspondence}
\label{AdS/CFT}
From the earlier lectures, we see that the symmetry group of CFT in $d$ dimensions is the same as the symmetry group of AdS spacetimes in $d+1$ dimensions. This hints toward the correspondence between $CFT_{d}$ and $AdS_{d+1}$ spacetime, i.e., ``{\it Einstein gravity with negative cosmological constant in $(d+1)$ dimensions is dual to $d$ dimensional CFT living at the boundary of $AdS_{d+1}$ spacetime.}'' A cartoon picture of this statement can be seen from Fig. \ref{AdS}, where we have to consider $AdS_{d+1}$ gravity in the bulk and $CFT_{d}$ at the boundary. Broadly, this duality has been termed as ``{\it gauge-gravity duality}'' which suggests ``{\it duality between gravity in $(d+1)$ dimensions and gauge theory in $d$ dimensions}''. 

First, we will provide the dictionary between the partition function of $AdS_{d+1}$ spacetime and $CFT_d$, where in the bulk we consider $\phi$ as a bulk scalar field, and then in sec. \ref{BBS}, we will discuss a broader version of this duality, including string theory. Let us consider the Einstein-scalar theory, which has the action given in the following form
\be \label{IAdS}
I[G,\phi]=\frac{1}{16 \pi G_N}\int d^{d+1}w\sqrt{G}\left[\mathcal{R}-2\Lambda+\frac{1}{2}\left(\nabla\phi\right)^{2}+\frac{1}{2}m^{2}\phi^{2}\right],
\ee
where $G$ is the bulk metric, $\phi$ is the bulk scalar field, $G_N$ being the Newton's constant and $\mathcal{R}$ is the Ricci scalar for $AdS_{d+1}$ spacetime. Solutions of the field EOMs appearing in (\ref{IAdS}) are given below with suitable boundary conditions for the metric and scalar field
\begin{eqnarray}
ds^{2} & = & G_{\alpha\beta}dw^{\alpha}dw^{\beta}=R^{2}\left[\frac{dz^{2}+dx^{\mu}dx^{\nu}\left[g_{\mu\nu}(x)+O(z)\right]}{z^{2}}\right]\ ,\label{eq:BCAdSmetric}\\
\hskip -0.1in \phi & = & \frac{z^{d-\Delta}}{2\Delta-d}\left[\phi_{b}(x)+O(z)\right]\ ,\nonumber 
\end{eqnarray}
 where $g$, $\phi_b$ are boundary quantities and $R$ is AdS length scale. According to AdS/CFT duality, partition functions of boundary and bulk theories are related by  \cite{Gubser:1998bc}\cite{Witten:1998qj}
\begin{equation}
Z[g_{\mu\nu},\phi_{b}]=\int_{{G\to g\atop \phi\to\phi_{b}}}\left[dG\right]\left[d\phi\right]e^{-I[G,\phi]} \, .\label{eq:AdSgeneratingfunctionmetric-i-g}
\end{equation}
 In the semiclassical limit, (\ref{eq:AdSgeneratingfunctionmetric-i}) reduces to
\begin{equation}
Z[g_{\mu\nu},\phi_{b}] \sim e^{-I[G,\phi]_{on-shell}}\, ,\label{eq:AdSgeneratingfunctionmetric-i}
\end{equation}
where $I[G,\phi]_{on-shell}$ is the on-shell action evaluated for the solution for metric and scalar field given in (\ref{eq:BCAdSmetric}). Therefore, by just computing the on-shell action in the bulk, we get the partition function of conformal field theory located at the boundary of $AdS_{d+1}$ spacetime. As we know, once we know the partition function, we can compute observables in our theory, such as correlation functions. This will be clearer when we do the explicit computation in sec. \ref{twoptAdS}.

\subsection{Observables and correlation functions}
\label{BBS}
According to AdS/CFT correspondence, every bulk field $\phi$ corresponds to an operator ${\cal O}$ in the boundary theory by the argument that the boundary value of $\phi$, i.e., $\phi_b$, acts as a source for the operator ${\cal O}$. Which can be stated by the following dictionary between the partition function of the bulk [if the bulk theory is string theory] and boundary theories
\bea 
& & Z_{s}[\phi_{b}]= \langle e^{\int d^4x \phi_{b} {\cal O}} \rangle_{CFT}=\int D \phi \, e^{- S[\phi]+\int d^d x \phi_b(x){\cal O}(x)},
 \label{eq:AdSgeneratingfunctionmetric-ii}
\eea
where $Z_{s}[\phi_{b}]$ is the string/supergravity partition function evaluated in $(d+1)$ dimensional bulk ($B_{d+1}$) with boundary conditions as mentioned earlier. The above relation is known as the GKPW relation \cite{Gubser:1998bc}\cite{Witten:1998qj}. In the classical supergravity (SUGRA) limit
\bea \label{Isugra}
& &  Z_{s}[\phi_{b}]= e^{- I_s[\phi]}|_{\partial B_{d+1}}.
\eea
Then $ n$-point function in the boundary theory is given by
\bea \label{npt}
& & \langle {\cal O}_1(x_1) {\cal O}_2(x_2) {\cal O}_3(x_3)........{\cal O}_n(x_n) \rangle=\frac{1}{Z[0]}\frac{\delta^n Z_{s}[\phi_{b}]}{\delta \phi_b(x_1)\delta \phi_b(x_2)\delta \phi_b(x_3).....\delta \phi_b(x_n)},
\eea
where $Z[0]$ is the normalization factor and we need to set $\phi_b \rightarrow 0$ at the end of calculation.
We can write (\ref{npt}) also as 
\bea \label{npt-a}
& & \langle {\cal O}_1(x_1) {\cal O}_2(x_2) {\cal O}_3(x_3)........{\cal O}_n(x_n) \rangle=\frac{\delta^n \log Z_{s}[\phi_{b}]}{\delta \phi_b(x_1)\delta \phi_b(x_2)\delta \phi_b(x_3).....\delta \phi_b(x_n)}.
\eea
If there is a gauge field ($A^a$) in the bulk with bounadry value $A^a_b$, then (\ref{eq:AdSgeneratingfunctionmetric-ii}) can be written as
\bea 
& & Z_{s}[A_{b}^a]= \langle e^{\int d^4x A_{b}^a J_a} \rangle_{CFT}.
 \label{eq:AdSgeneratingfunctionmetric-iii}
\eea
In the SUGRA limit, $Z_{s}[A_{b}^a] = e^{-I_s[A_{b}^a]}$ and we can compute $n$ point function similar to (\ref{npt}) as follows
\bea \label{npt-i}
& & \langle J_a^1(x_1) J_a^2(x_2) J_a^3(x_3)........J_a^n(x_n) \rangle=\frac{\delta^n \log Z_{s}[A_{b}^a]}{\delta A_{b}^a(x_1)\delta A_{b}^a(x_2)\delta A_{b}^a(x_3).....\delta A_{b}^a(x_n)}.
\eea
We can generalize this to the stress tensor correlator at the boundary as
\bea \label{npt-ii}
& & \hskip -0.5in \langle T_{\mu_1 \nu_1}(x_1) T_{\mu_2 \nu_2}(x_2) T_{\mu_3 \nu_3}(x_3)........T_{\mu_n \nu_n}(x_n) \rangle=\frac{\delta^n \log Z_{s}[g_{\mu \nu}]}{\delta g_{\mu_1 \nu_1}(x_1)\delta g_{\mu_2 \nu_2}(x_2)\delta g_{\mu_3 \nu_3}(x_3).....\delta g_{\mu_n \nu_n}(x_n)}.\nn \\
\eea
where $g$ is the boundary value of the bulk metric $G$. We should mention that the partition function should be regulated even in the semiclassical limit using holographic renormalization by adding suitable counterterms \cite{Balasubramanian:1999re}\cite{Skenderis:2002wp}. See \cite{Yadav:2021hmy} for holographic renormalization of eleven-dimensional SUGRA action appearing in $M$ theory and \cite{Nanda:2025tid} for counterterms in the context of de Sitter space as a bulk theory. \\
{{\bf Exercise}: Derive the form of brown york stress tensor as given in \cite{Balasubramanian:1999re}. Use it to obtain the stress tensor for $AdS_3$ background and, from the trace, derive the central charge. Finally, show that the stress tensor is traceless in $AdS_4$ spacetime.} 

One evidence of AdS/CFT correspondence comes from the boundary stress tensor correlators computed from the bulk and boundary perspectives. For example, connected correlators of stress tensor computed from the bulk in the semiclassical limit ($l_p << L$) scale as (with $l_p$ being the Plank length) 
\bea \label{npt-Tmunu}
& & \hskip -0.5in \langle T_{\mu_1 \nu_1}(x_1) T_{\mu_2 \nu_2}(x_2) T_{\mu_3 \nu_3}(x_3)........T_{\mu_n \nu_n}(x_n) \rangle \sim \left(\frac{R}{l_p} \right)^{d-1}\nn \\
\eea
From the CFT calculations, for $T_{\mu \nu} \sim \frac{N}{\lambda} {\rm Tr}\left(\partial_\mu \Phi \partial_\nu \Phi\right)$, in the large $N$ limit 
\bea \label{npt-Tmunu-i}
& & \hskip -0.5in \langle T_{\mu_1 \nu_1}(x_1) T_{\mu_2 \nu_2}(x_2) T_{\mu_3 \nu_3}(x_3)........T_{\mu_n \nu_n}(x_n) \rangle \sim N^2.\nn \\
\eea
Hence, the large $N$ limit of CFT is related to the semiclassical limit of gravity in the AdS, provided $N^2 \sim \left(\frac{R}{l_p} \right)^{d-1}$.

\subsection{Basics of string theory, $Dp$-branes, and SUSY}
{\bf String theory}: String theory is a candidate theory where we can see the unification of all forces in nature, such as strong, weak, electromagnetic, and gravitational interactions. In string theory, a fundamental object is a one-dimensional ``string'' which moves throughout the spacetime and interacts with other strings. Particles are excitations of the string. There are two kinds of Strings
\begin{itemize}
    \item Open strings 
    \item Closed strings
\end{itemize}
Open strings satisfy certain boundary conditions. Let us consider an open string with ends at $\sigma=0$ and $\sigma=\pi$ when it propagates in spacetime, we get a worldsheet made up of $(\tau, \sigma)$ coordinates. Let us denote this embedding as $X^\mu(\tau, \sigma)$ in the ambient spacetime then the boundary conditions are
\begin{itemize}
\item {Neumann boundary condition}: 
\begin{equation}
\left(\cfrac{\partial X^\mu(\tau,\sigma)}{\partial\sigma}\right)_{\sigma=0} = \left(\cfrac{\partial X^\mu(\tau,\sigma)}{\partial\sigma}\right)_{\sigma=\pi}.
\end{equation}

\item Dirichlet boundary condition: 
\begin{equation}
X^\mu(\tau,\sigma)\vert_{\sigma=0} = X^\mu(\tau,\sigma)\vert_{\sigma=\pi}.
\end{equation}
\end{itemize}
\begin{figure}[h]
\begin{center}
\includegraphics[width=0.8\textwidth]{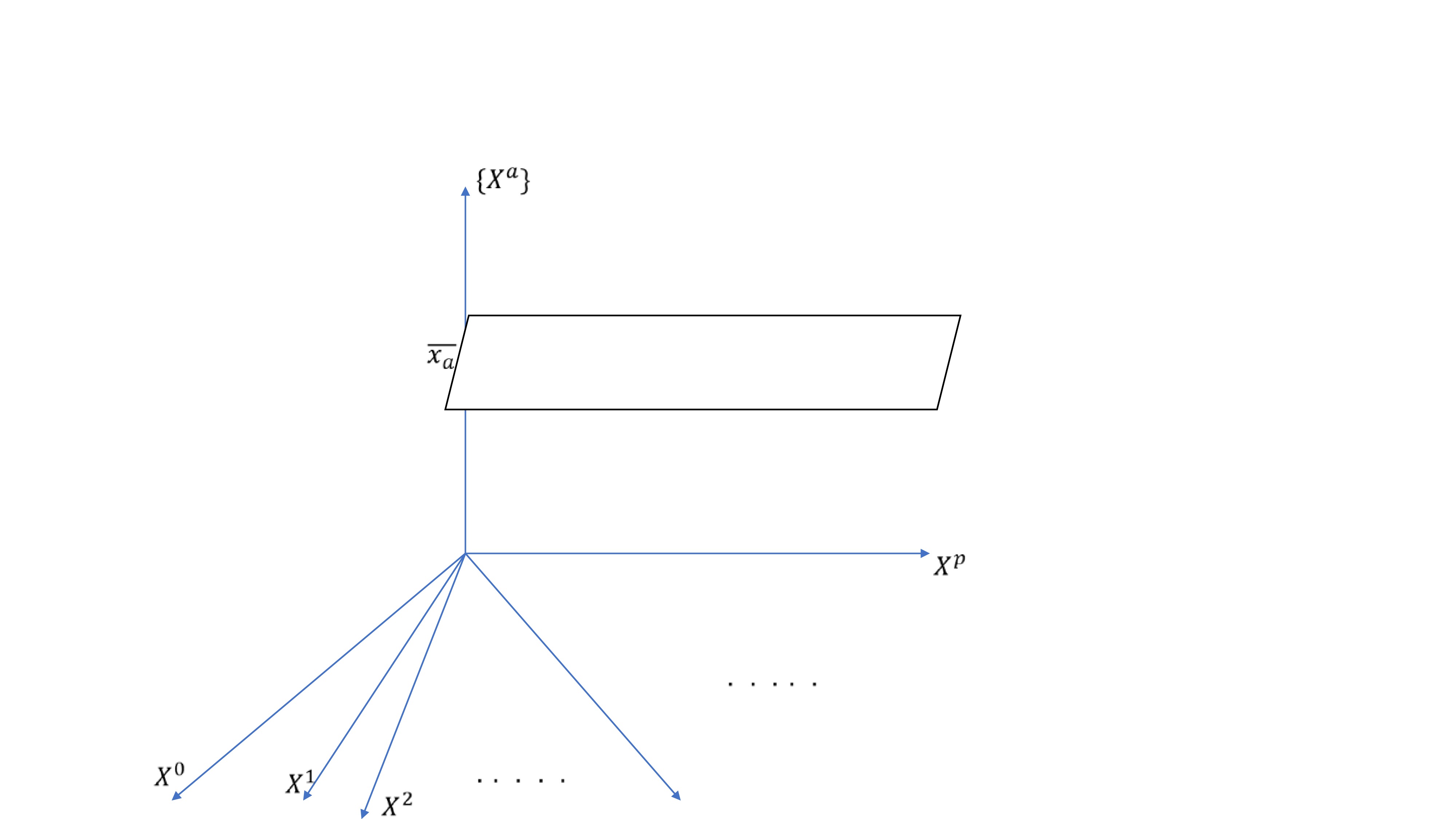}
\end{center}
\caption{In this figure, we have $Dp$-brane located at $X^a=\overline{x}_a$.}
\label{Dp-brane}
\end{figure}
{\bf $Dp$-branes}: $Dp$-branes are extended objects in string theory with $p$ spatial dimensions. In the ambient $d+1$ dimensional spacetime, coordinates along the $Dp$-branes are $X^0, X^1,.....,X^p$ and coordinates orthogonal to the $Dp$-branes are $X^a$ where, $a = p+1,p+2, ......,d$. Orthogonal coordinates provide the location of $Dp$-branes, see Fig. \ref{Dp-brane} for example. As we discussed earlier, when a one-dimensional string propagates in spacetime, it draws a two-dimensional worldsheet. For $Dp$-branes, we get $(p+1)$ dimensional world volume.

{\bf Open Strings between parallel $Dp$-branes.}
\begin{figure}[h]
    \centering
    \includegraphics[width = 15cm]{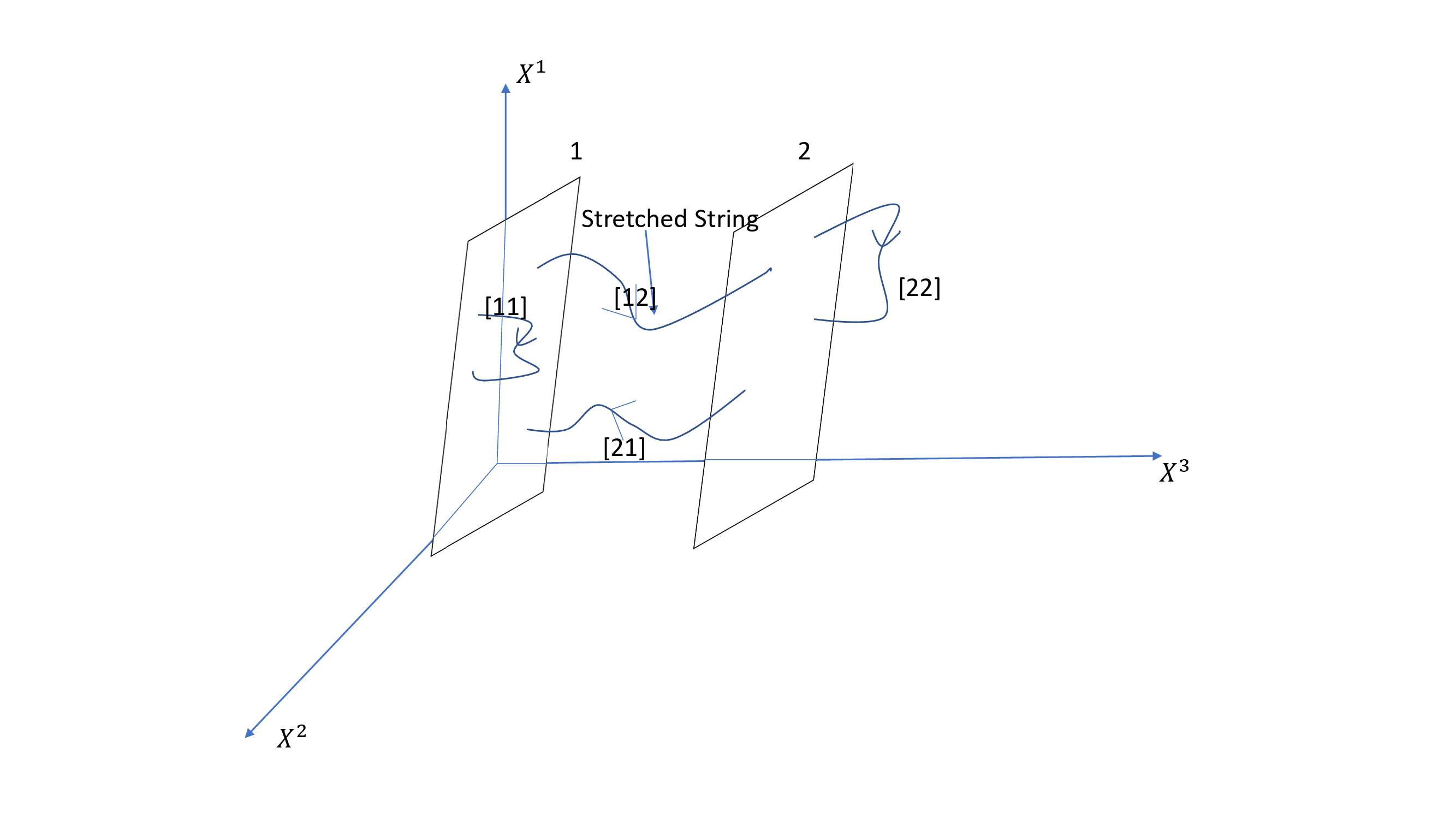}
    \caption{In this figure, we have a cartoon picture of open strings attached between two $Dp$-branes.}
    \label{pDp}
\end{figure}
Strings attached between two $Dp$ branes can give massive modes if the branes are separated; otherwise, we will get massless modes. For two $Dp$-branes, we have four sectors as shown in Fig. \ref{pDp}.

 If we take $N$ coincident $Dp$-branes then each sector is represented by $[ij]$, where $i,j = 1,2,......,N$. Overall, we have $N^2$ sectors, which gives $U(N)$ Yang-Mills theory on the world volume of $N$ coincident $Dp$-branes. This is what we wanted to show from this brief review because this has been used in Maldacena's conjecture, about which we are going to discuss in sec. \ref{MC}.

{\bf Supersymmetry (SUSY)}: Here we want to give a flavor of what supersymmetry is from \cite{Green:2012oqa}. For simplicity, we will discuss worldsheet supersymmetry, which relates spacetime coordinates $X^\mu(\tau,\sigma)$ to the fermionic partners $\psi^\mu(\tau,\sigma)$, which are two-component world sheet spinors. The action for this system is
\bea \label{S-susy}
& & S=-\frac{1}{2\pi}\int d^2\sigma \left(\partial_\alpha X^\mu \partial^\alpha X_\mu - i\overline{\psi}^\mu \rho^\alpha \partial_\alpha \psi_\mu \right),
\eea
where $\rho^\alpha$ are two-dimensional Dirac matrices defined as
\bea
& & \rho^0=\begin{pmatrix}
0 & -i \\
i & 0
\end{pmatrix}, \ \ \  \ \ \ \ \ \rho^1=\begin{pmatrix}
0 & i \\
i & 0
\end{pmatrix}.
\eea
Two-dimensional Dirac matrices follow
\bea
& & \{\rho^\alpha,\rho^\beta \}=-2 \, \eta^{\alpha \beta},
\eea
where $\eta_{\alpha \beta}$ is flat two dimensional metric. Under the following supersymmetric transformations
\bea \label{SUSY-T}
&& \hskip -0.1in \delta X^\mu =\overline{\epsilon} \, \psi^\mu, \nn \\
& & \delta \psi^\mu =- i \rho^\alpha (\partial_\alpha X^\mu)\epsilon \, ,
\eea
The action (\ref{S-susy}) remains invariant. From the SUSY transformations (\ref{SUSY-T}), we can clearly see that the variation of bosonic degrees of freedom is proportional to the fermionic degrees of freedom and vice versa. If we have supersymmetric theories, then we will have terms for the bosons and their supersymmetric partners in the action.

Now we proceed to discuss Maldacena's conjecture in the next section, where readers can see all these terms which we just discussed.

\subsection{Maldacena's conjecture} \label{MC}
Originally AdS/CFT was proposed by Maldacena in his seminal paper in 1997 \cite{Maldacena:1997re}. Maldacena proposed a duality between $\mathcal{N}=4$ SYM (supersymmetric Yang-Mills) theory in four dimensions and type IIB supergravity on $AdS_5 \times S^5$. Here we will discuss this duality.

The central idea is to analyze the low-energy dynamics of $D$-branes configuration through the perspective of open and closed strings. To clarify this, let us briefly recall the standard example that leads to the AdS/CFT correspondence \cite{Maldacena:1997re}. Take $N$ coincident $D3$-branes in type IIB string theory embedded in ten-dimensional Minkowski spacetime. Closed strings moving in the bulk of this spacetime can interact with the $D3$-branes, and such interactions admit two descriptions:

(a) A $D3$-brane may be viewed as a submanifold that serves as an endpoint for open strings. In this picture, the interaction of a closed string with the brane can be understood as the closed loop splitting into an open string whose endpoints are anchored on the D3-brane.

(b) $D3$-branes can also be understood as solitonic solutions within closed string theory; equivalently, they act as sources that generate a nontrivial curved spacetime in which closed strings propagate.

\begin{figure}
\begin{centering}
\includegraphics[width=0.6\textwidth]{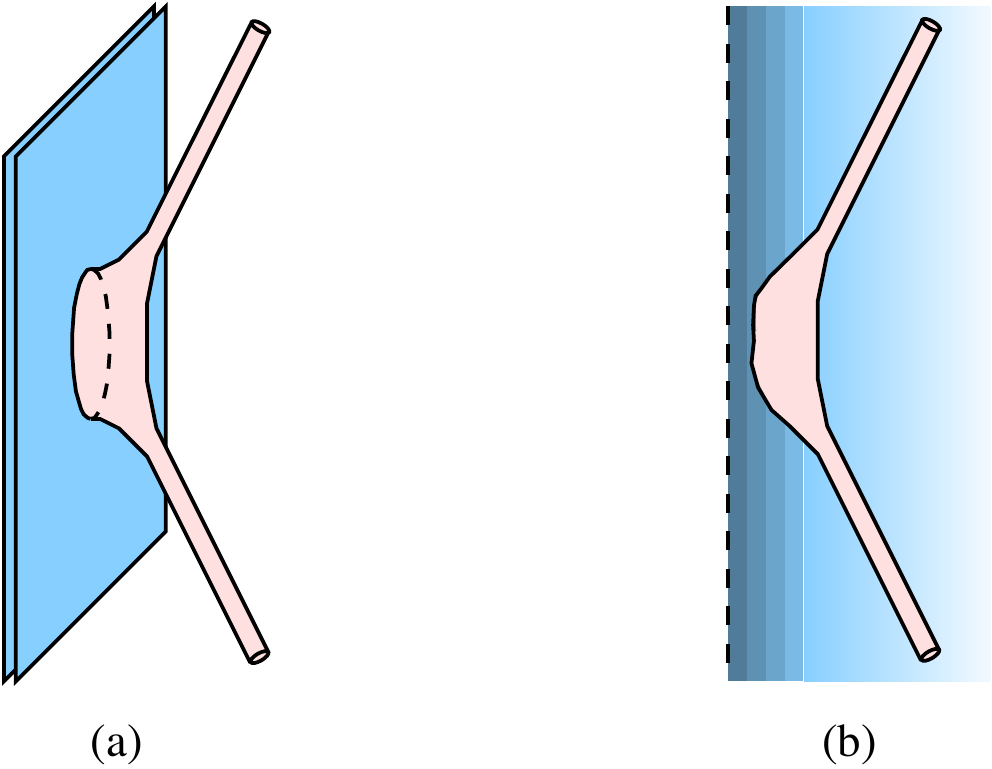}
\par\end{centering}

\protect\caption{\label{fig:OpenClosedDuality}(a) (a) Scattering of a closed string from branes in flat spacetime. (b) Propagation of a closed string in a curved geometric background. Figure adapted from \cite{Penedones:2016voo}.}

\end{figure}

The two descriptions are illustrated in Fig. \ref{fig:OpenClosedDuality}, and their equivalence is known as open/closed duality. The AdS/CFT correspondence emerges as the low-energy limit of this duality. To realize this limit, one sends the string length $\ell_{s}\to 0$ while keeping fixed the string coupling $g_{s}$, the number of branes $N$, and the characteristic energy scale.
 \par
In description (a), the low-energy spectrum splits into two independent sectors:\\
$[i]$ massless closed strings propagating in ten-dimensional Minkowski spacetime, and\\
$[ii]$ massless open strings ending on the D3-branes, which at low energies are effectively described by $\mathcal{N}=4$ supersymmetric Yang-Mills theory with gauge group $SU(N)$.
  \par
In description (b), the low-energy dynamics also separates into two distinct sectors:\\
$[i]$ massless closed strings propagating in the curved background
\begin{equation} \label{10D}
ds^{2}=\frac{1}{\sqrt{H(r)}},\eta_{\mu\nu}dx^{\mu}dx^{\nu}+\sqrt{H(r)}\big(dr^{2}+r^{2}d\Omega_{5}^{2}\big),
\end{equation}
where $\eta_{\mu\nu}$ is the four-dimensional Minkowski metric along the branes and 
\be
H(r)=1+\frac{L^{4}}{r^{4}}\ ,\qquad L^{4}=4\pi g_{s}N\ell_{s}^{4}\ .
\ee
At first sight, sending $\ell_{s}\to 0$ simply recovers ten-dimensional Minkowski space.\\ 
 $[ii]$ However, in the near-horizon region ($r\to 0$), $H(r)$ reduces to $L^{4}/r^{4}$, and the geometry (\ref{10D}) becomes
\begin{equation} \label{10D-i}
ds^{2}=\frac{r^{2}}{L^{2}}\eta_{\mu\nu}dx^{\mu}dx^{\nu}+\frac{L^{2}}{r^{2}}dr^{2}+L^{2}d\Omega_{5}^{2},
\end{equation}
which is precisely the metric of $AdS_{5}\times S^{5}$. This is made explicit by defining the coordinate $z=L^{2}/r$, giving
\be
ds^{2}=\frac{ L^{2}}{z^{2}}\left[dz^{2}+\eta_{\mu\nu}dx^{\mu}dx^{\nu}\right]+L^{2}d\Omega_{5}^{2}\ ,
\ee
the standard Poincaré patch of $AdS_{5}\times S^{5}$, with common radius $L$ for both $AdS_{5}$ and $S^{5}$.

Thus, description (b) again yields two decoupled low-energy sectors:\\
$[i]$ massless closed strings in flat ten dimensions, and\\
$[ii]$ full type IIB string theory on $AdS_{5}\times S^{5}$.

This observation led Maldacena to propose the celebrated conjecture that
\begin{center} \boxed{ \begin{tabular}{ccc} $\mathcal{N}=4$ SYM theory with gauge group $SU(N)$ & \multirow{2}{*}{$\Leftrightarrow$} & Type IIB string theory on $AdS_{5}\times S^{5}$ \\[6pt] $g_{\text{YM}}^{2}=4\pi g_{s}$ & & $\dfrac{R^{4}}{\ell_{s}^{4}} = g_{\text{YM}}^{2} N \equiv \lambda$ \\ \end{tabular} } \end{center}

$\mathcal{N}=4$ SYM remains conformal for arbitrary $N$ and coupling $g_{YM}^{2}$. The field content consists of the gauge field with strength tensor
\be
F_{\mu\nu}=\partial_{\mu}A_{\nu}-\partial_{\nu}A_{\mu}-i\left[A_{\mu},A_{\nu}\right]\ ,
\ee
six scalar fields $\Phi^{m}$, and four Weyl fermions $\Psi^{a}$, all transforming in the adjoint representation of $SU(N)$. The Lagrangian takes the form
\begin{align}
&{\cal L}=\frac{1}{g_{YM}^{2}}{\rm Tr}\left[\frac{1}{4}F^{\mu\nu}F_{\mu\nu}+\frac{1}{2}\left(D^{\mu}\Phi^{m}\right)^{2}+\bar{\Psi}^{a}\sigma^{\mu}D_{\mu}\Psi_{a}\right.\\
&\qquad\qquad\left.-\frac{1}{4}\left[\Phi^{m},\Phi^{n}\right]^{2}-C_{m}^{ab}\Psi_{a}\left[\Phi^{m},\Psi_{b}\right]-\bar{C}_{mab}\bar{\Psi}^{a}\left[\Phi^{m},\bar{\Psi}^{b}\right]\right]\ , \nonumber
\end{align}
where $D{\mu}$ denotes the gauge covariant derivative, and $C_{m}^{ab}, \bar{C}_{mab}$ are constants fixed by the global $SO(6)\simeq SU(4)$ symmetry.
Great! We found the duality between $\mathcal{N}=4$ SYM theory in four dimensions and type $IIB$ SUGRA on $AdS_5 \times S^5$ background. Now we will match the symmetries on both sides of duality.
\begin{itemize}
\item $AdS_5 \times S^5$ has the symmetry group as $SO(2,4)$ coming from $AdS_5$ and $SO(6)$ coming from $S^5$.

\item $\mathcal{N}=4$ SYM theory in four dimensions is a type of conformal field theory with supersymmetry so conformal group is $SO(2,4)$. Further, Lagrangian of this theory contains six scalars $\Phi^m$ and four fermions $\Psi^a$ which are related via a global $SU(4) \simeq SO(6)$ $R$-symmetry.
\begin{itemize}
\item For scalars: $\Phi^m \rightarrow R^m_n \, \Phi^n$ where $R \in SO(6)$.
\item For fermions: $\Psi^a \rightarrow U^a_b \, \Psi^b$ where $U \in SU(4)$.
\item SUSY relates $\Phi^m$ with $\Psi^a$ and hence $SU(4) \simeq SO(6)$.
\end{itemize}

\end{itemize}
Amazing! We found perfect matching of symmetries on both sides. Further, one can show that both side also have 32 SUSYs, which manifest themselves as Killing spinors in $AdS_5 \times S^5$ and as superconformal algebra on CFT side.

{\bf Useful remarks:}
\begin{itemize}
\item {\bf Low energy limit}: $l_s \rightarrow 0$ because mass of the string scale as $m \propto  \frac{1}{l_s}$. Therefore in $l_s \rightarrow 0$ limit we can ignore massive strings and we will be left with massless modes. The reason is as follows. Excitations with energies $E>> \frac{1}{l_s}$ can excite full stringy modes while those with energy $E<< \frac{1}{l_s}$ only massless modes survive. This is also known as SUGRA limit.

\item {\bf Validity of SUGRA limit}: In order to trust the gravity solution (to suppress the string corrections to the geometry), we need to keep $L$ large which implies 't Hooft coupling $\lambda>>1$. We can see this from here
\bea
& & \frac{L}{l_s}=\left(4 \pi g_s N \right)^{1/4} \sim \lambda^{1/4}.
\eea
Consider the general action with stringy corrections of the form
\bea
& & S =\frac{1}{16 \pi G_N}\int d^{10x}\sqrt{-g}\left(R+l_s^2 R^2+l_s^4 R^4+....\right).
\eea
Ricci scalar scale as: $R \sim \frac{1}{L^2}$ which implies that when $L$ is large, Ricci scalar is small. We obtain $l_s^2 R=\frac{l_s^2}{L^2}$. Hence higher curvature corrections are small if $\frac{l_s^2}{L^2}<<1$ or $L>>l_s$. Classical SUGRA is the low energy limit of string theory.  

\item {\bf Suppression of quantum corrections}: In string theory, loop corrections (quantum corrections) come from string worldsheet diagram with handle (genus $g\geq 1$). Each extra loop or handle contributes and extra factor of $g_s^2$. This implies that quantum corrections are suppressed by powers of string coupling when $g_s << 1$. To see this explicitly, consider the action
\bea
& & S \sim \frac{1}{\kappa_{10}^2}\int d^{10}x \sqrt{-g}R = \frac{1}{g_s^2 l_s^8}\int d^{10}x \sqrt{-g}R
\eea
Therefore, when $g_s$ is very small then  $\frac{1}{g_s^2}$ is large and hence quantum corrections are suppressed. Therefore in $g_s<<1$ limit, classical (tree-level) SUGRA is reliable.

\end{itemize}

\subsection{Finite temperature holography \label{sub:Finite-Temperature}}
Thermal state on the field theory side requires a black hole on the gravity dual side. To be more precise, an AdS black hole is dual to a thermal state on the field theory side. Consider Schwarzschild $AdS_5 \times S^5$ black hole with the following metric
\bea \label{SAdS}
& & ds^2=-g(r) dt^2+\frac{dr^2}{g(r)}+r^2 d\Omega_3^2 + L^2 d\Omega_5^2,
\eea
where 
\bea
& & g(r)=\frac{r^2}{L^2}+1-\frac{r_0^2}{r^2} ; \ \ \ \ r_0^2=\frac{8 G M}{3 \pi}=r_+^2\left(\frac{r_+^2}{L^2}+1\right).
\eea
Here $G$ is the five-dimensional Newton's constant. In the large black hole limit, $r_+ >> L$, the above metric (\ref{SAdS}) reduces to the following form
\bea
& & ds^2=\frac{r^2}{l^2}\left[- \left(1-\frac{r_+^4}{r^4} \right)dt^2+\sum_{i=1}^3 dx_i dx^i\right]+\frac{l^2}{r^2}\left(1-\frac{r_+^4}{r^4} \right)^{-1} dr^2+L^2 d\Omega_5^2.
\eea
Large black hole on the AdS side corresponds to the hot plasma of gauge theory degrees of freedom at the Hawking temperature. Hawking temperature for the solution (\ref{SAdS}) is
\bea
& & T=\frac{g'(r_+)}{4\pi}=\frac{2 r_+^2+ L^2}{2\pi r_+ L^2}.
\eea 
The behaviour of small ($r_+ <<L$) and large ($r_+ >> L$) AdS black holes are: $T \sim \frac{1}{r_+}$ and $T \sim \frac{r_+}{L^2}$ respectively. The Latter corresponds to a high-temperature thermal state in CFT. One can see the analogy of phase transition from the confined phase to the deconfined phase as phase transition from a small black hole to a large black hole, see \cite{Witten:1998zw}\cite{Hubeny:2014bla} for details.

\section{Recent developments}
%Explicit examples: consistency checks}
\label{C-checks}
We will discuss three consistency checks for the AdS/CFT correspondence based on recent developments in this field.
\subsection{Holographic entanglement entropy (HEE)} 
Here, our aim is to calculate entanglement entropy in boundary field theory using the $AdS/CFT$ correspondence. To do so, we will start with a review of how to calculate entanglement entropy of quantum mechanical systems, EE in QFT using the replica trick, and entanglement entropy in CFT using the Ryu-Takayangai formula. Finally, we will use the RT formula to calculate the entanglement entropy of an interval of CFT in the vacuum state in sec. \ref{EE-ads3cft2}.\\
{\bf Entanglement entropy in quantum mechanics (QM)} :
To compute the entanglement entropy for quantum mechanical systems, consider a bipartite system whose state is denoted by $|\psi \rangle_{AB}$. The density matrix of this system is defined as:
\begin{equation}
\label{dm-qm}
\rho_{AB} = {|\psi \rangle}_{AB}  {\langle \psi|}_{AB}.
\end{equation}
Entanglement between subsystems $A$ and $B$ is measured by the von-Neumann entropy defined as follows
\begin{eqnarray}
\label{vN-entropy-defn}
S_{\rm EE}=- {\rm Tr} \left(\rho_A {\rm ln}\rho_A \right),
\end{eqnarray}
where $\rho_{A}$ is the reduced density matrix obtained by tracing over $B$
\begin{equation}
\label{red-dm-A}
\rho_{A}={\rm Tr}_B\left(\rho_{AB}\right)={\rm Tr}_B\left({|\psi \rangle}_{AB}  {\langle \psi|}_{AB}\right) = {|\psi \rangle}_A  {\langle \psi|}_A.
\end{equation}

{\bf Entanglement entropy in quantum field theory (QFT)}: In quantum field theory (QFT), computing entanglement entropy is subtle because the Hilbert space does not naturally factorize into independent subsystems. To address this, one typically uses the replica trick. The starting point is the Rényi entropy, defined as
\begin{eqnarray}
\label{Renyi-entropy}
S_{A}^{(n)}=\frac{1}{1-n} \log\left({\rm Tr} \rho_A^n\right),
\end{eqnarray}
where $\rho_A$ is the reduced density matrix of the subsystem $A$. The entanglement entropy (or von Neumann entropy) is then obtained in the limit: $S_{\rm EE}= {\lim_{n\rightarrow 1}}S_{A}^{(n)}$. 
In practice, the calculation proceeds by considering $n$ replicated copies of the subsystem $A$. The operation ${\rm Tr} \rho_A^n$ corresponds to cyclically gluing these $n$ replicas along the subsystem $A$, producing an $n$-sheeted Riemann surface. One finds that
\begin{eqnarray}
{\rm Tr}\rho_A^n=\frac{Z_n(A)}{Z_1^n}\, ,
\end{eqnarray}
with $Z_n(A)$ the partition function on this branched $n$-sheeted geometry, and $Z_1$ the partition function of the original theory. Thus, the Rényi entropy in QFT can be expressed as
\begin{eqnarray}
\label{Renyi-entropy-QFT}
S_{A}^{(n)}=\frac{1}{1-n} \log\left(\frac{Z_n(A)}{Z_1^n}\right).
\end{eqnarray}
Taking the analytic continuation $n \rightarrow 1$ then yields the von Neumann entropy.

{\bf Entanglement entropy from holography}: In the earlier discussion, we noted that entanglement entropy in QFT can be obtained once the partition function on the $n$-sheeted geometry is known. However, computing this partition function directly is often highly nontrivial. The AdS/CFT correspondence offers a more tractable approach through the Ryu-Takayanagi (RT) prescription \cite{Ryu:2006bv}, which relates the entanglement entropy of a boundary CFT region to a geometric quantity in the dual gravitational theory.
\begin{figure}[h]
\begin{center}
\includegraphics[width=0.4\textwidth]{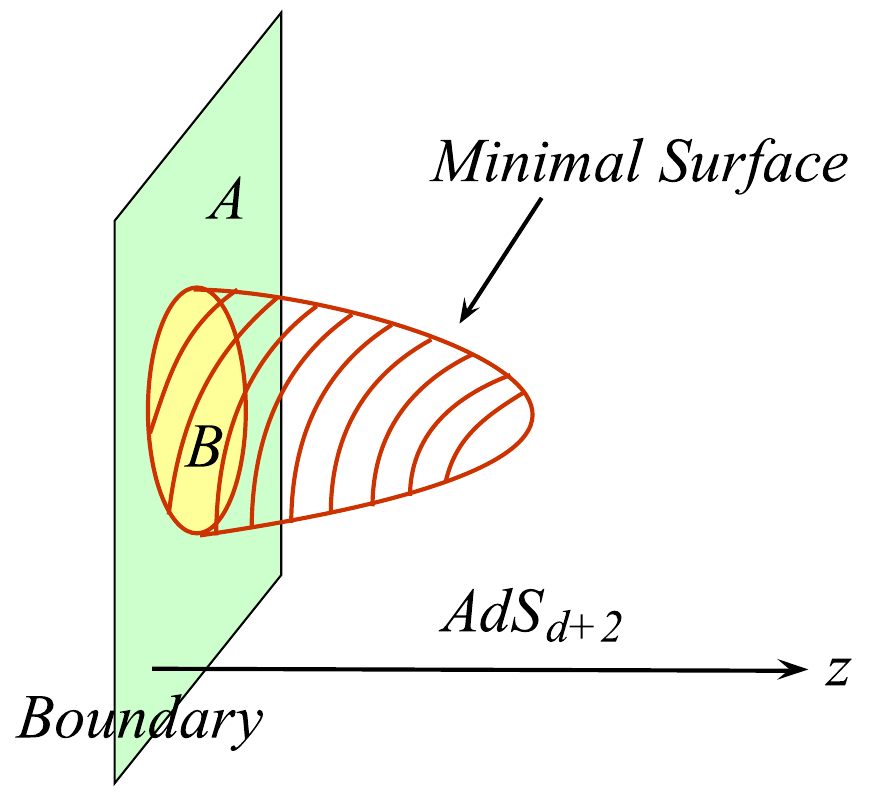}
\end{center}
\caption{ Entanglement entropy from $AdS/CFT$. This figure is taken from \cite{Nishioka:2009un}.}
\label{RT}
\end{figure}
Consider the $AdS_{d+2}/CFT_{d+1}$ correspondence with bulk spacetime ${\cal M}_{d+2}$. Let $B$ be a chosen subsystem on the CFT. The RT proposal can be summarized as follows. Consider a co-dimension two surface $\epsilon_B$ anchored on the boundary of subergion $B$ ($\partial B$) in the bulk ${\cal M}_{d+2}$. Out of many surfaces, we have to consider the one that satisfies the homology constraint, i.e., $\epsilon_B$ is smoothly retractable to the boundary region $A$. Among all such surfaces, select the one with minimal area. See Fig. \ref{RT} for the pictorial description of Ryu-Takayanagi prescription. The entanglement entropy is then given by
\begin{eqnarray}
\label{HEE-RT-defn}
S_B=\frac{{\rm min} \left({\rm Area}(\epsilon_B)\right)}{4 \, G_{d+2}}.
\end{eqnarray}

The RT prescription applies to static (time-independent) backgrounds. For time-dependent spacetimes, one must instead use the Hubeny-Rangamani-Takayanagi (HRT) prescription \cite{Hubeny:2007xt}. 
%Beyond the classical RT/HRT formulas, quantum corrections to all orders in $\hslash$ were introduced in \cite{Engelhardt:2014gca}. In this framework, one extremizes the generalized entropy rather than the area alone, and the corresponding extremal surfaces are known as quantum extremal surfaces (QES). If multiple QES exist, the one with minimal generalized entropy is chosen.

\subsubsection{Holographic entanglement entropy in $AdS_3/CFT_2$ correspondence}
\label{EE-ads3cft2}
Here we will calculate the entanglement entropy for a single interval in a CFT$_2$ on $\mathbb{R}^{1,1}$. Consider an interval with length $2a$ centred around the origin on a spacelike slice at $t=0$ such that 
\bea
& & A= \{x \in \mathbb{R}| -a < x < a \}; \ \ \ \partial A=\{-a,  a\}
\eea
The holographic dual of CFT$_2$ vacuum is given by the Poincaré patch of $AdS_3$ spacetime. The metric of $AdS_3$ spacetime in the Poincaré patch is given as
\begin{equation}
ds^2 = \frac{L^2}{z^2} (-dt^2 + dx^2 + dz^2).
\end{equation}
Applying the Ryu-Takayanagi prescription \cite{Ryu:2006bv} with the extremal surface parametrized as $z(x)$, the holographic entanglement entropy is given by
\begin{equation} \label{HEE-AdS3}
S_A = \frac{1}{4G_3} \int_{-a}^{a} dx \, \frac{L}{z(x)}\sqrt{1+\left(\frac{dz(x)}{dx}\right)^2}=\frac{1}{4G_3} \int_{-a}^{a} dx \, {\cal L}(z(x),z'(x)),
\end{equation}
where $z'(x)=\frac{dz(x)}{dx}$. Extremization of (\ref{HEE-AdS3}) requires solving the equation of motion for $z(x)$ as
\bea \label{EOM-z(x)}
& & \frac{\partial {\cal L}(z(x),z'(x))}{\partial z(x)}-\frac{d}{dx}\left(\frac{\partial {\cal L}(z(x),z'(x))}{\partial z'(x)}\right)=0, \nn \\
& & {\rm implying} \ \ \ \  \ \ \  z(x) z''(x)+ z'(x)^2+1=0.
\eea
Solution of (\ref{EOM-z(x)}) is a semicircle around the origin in $z-x$ plane: $z(x)=\sqrt{a^2-x^2}$ . Substituting this solution into (\ref{HEE-AdS3}), we obtain
\bea
& & S_A=\frac{L}{4 G_3}\int_{-a}^{a} dx \, \frac{a}{a^2-x^2}=\frac{L}{2 G_3}\int_{0}^{a} dx \, \frac{a}{a^2-x^2}.
\eea
We can evaluate the $x$ integral by using $x=a \cos \theta$ which leads to
\bea
& & I=\int_{\epsilon}^{a} dx \, \frac{a}{a^2-x^2}=\int_{\epsilon/a}^{\pi/2} d\theta \, \frac{1}{\sin \theta}=\log\left(\frac{2 a}{\epsilon}\right).
\eea
Therefore, entanglement entropy is 
\bea
& & S_A=\frac{c}{3} \log\left(\frac{2 a}{\epsilon}\right),
\eea
where $c=\frac{3L}{2 G_3}$ is central charge of two dimensional CFT. This matches precisely the CFT$_2$ result for entanglement entropy as given in \cite{Calabrese:2004eu}\cite{Calabrese:2009qy}. See \cite{Nishioka:2009un}\cite{Kibe:2021gtw} for a review on holographic entanglement entropy.

\subsection{Holographic complexity}
The notion of quantum circuit complexity provides a measure of how challenging it is to prepare a desired target state $|\Psi_T\rangle$ from a chosen reference state $|\Psi_R\rangle$. Concretely, one may express the transformation as
\bea
& & | \Psi_T \rangle = U_{TR} |\Psi_R \rangle = g_n \ g_{n-1} \ g_{n-2}.........g_2 \ g_1 |\Psi_R \rangle \, ,
\eea
where the unitary $U_{TR}$ is built from a sequence of elementary gates ${g_i}$. The circuit complexity of $|\Psi_T\rangle$ is then defined as the minimal number of gates required to implement such a unitary. Importantly, in quantum systems the complexity continues to increase even after thermalization, a behavior that reflects the linear growth of the black hole interior on the gravitational side \cite{Hartman:2013qma}.\par

In holography, the complexity of the boundary theory is calculated using quantum information theoretic tools, and several competing proposals have been put forward. One such proposal is the “complexity = volume” (CV) conjecture \cite{Susskind:2014rva,Stanford:2014jda,Susskind:2014jwa,Roberts:2014isa,Susskind:2014moa}, where the complexity is associated with the volume of a codimension-one extremal hypersurface in the bulk spacetime. Other well-studied proposals include “complexity = action” \cite{Brown:2015bva}, “complexity = spacetime volume” \cite{Couch:2016exn}, and the more recent “complexity = anything” perspective \cite{Belin:2021bga,Belin:2022xmt}. Reviews and comparisons of these approaches can be found in \cite{Chapman:2021jbh,Myers:2024vve,Baiguera:2025dkc}.

Focusing on the CV proposal, studies of eternal black holes: which are dual to thermofield double states, together with the ER=EPR conjecture \cite{Maldacena:2013xja}, indicate that the spatial volume of the Einstein-Rosen bridge grows linearly with boundary time. This growth is the analogue of the linear increase in circuit complexity of the dual quantum state \cite{Susskind:2014rva,Stanford:2014jda,Susskind:2014jwa,Roberts:2014isa,Susskind:2014moa}. In this case, the complexity is estimated as
\be \label{CV}
C(t) \sim {{\rm Vol}(\Sigma_t)\over G_N R}\,,
\ee
where $\Sigma_t$ is an extremal codimension-one spacelike slice anchored at boundary time $t$, $R$ is the AdS curvature radius, and $G_N$ the Newton constant.

In static spacetimes, the dominant contribution to ${\rm Vol}(\Sigma_t)$ arises from the near boundary region. Introducing a cutoff $\epsilon$ in a geometry with $d_i$ spatial boundary directions yields the scaling relation for the complexity as
\bea
& & C(t)\propto {R^{d_i+1}\over G_{d_i+2}\,R}\, {V_{d_i}\over\epsilon^{d_i}}
\equiv N_{dof}\, V_{d_i}\Lambda_{_{UV}}^{d_i}\,.
\eea
 This expression shows that holographic complexity is proportional both to the number of effective degrees of freedom of the boundary theory and to the regulated spatial volume, expressed in terms of the UV cutoff scale.
 
We will discuss one example of holographic complexity which connects holography, cosmology, and quantum information. This is based on \cite{Narayan:2024fcp} where we studied the complexity of cosmologies with spacelike Big-Bang/Crunch singularities \cite{Das:2006dz}\cite{Awad:2007fj}. \\
~\\
{\bf Cosmological backgrounds}
\begin{itemize}
\item Consider the general ansatz for a $D=d_i+2$ dimensional gravity background and its two dimensional reduction 
\bea\label{dimredAnsatz}
&& ds_D^2 = g_{\mu \nu}^{(2)} dx^\mu dx^\nu +\phi^{2/d_i}d\sigma_{d_i}^2
= \frac{e^f}{\phi^{(d_i-1)/d_i}}(-dt^2+dr^2)+\phi^{2/d_i}d\sigma_{d_i}^2\,,\nn\\
[1mm]
&& ds^2 = g_{\mu\nu} dx^\mu dx^\nu = e^f\,(-dt^2+dr^2)\,,\qquad 
g_{\mu\nu} = \phi^{(d_i-1)/d_i} g_{\mu \nu}^{(2)}\,.
\eea
%\vspace{0.5mm}
\item  In the vicinity of
the singularity: $\phi=t^kr^m\,, e^f=t^ar^b\,,  e^\Psi=t^\al r^\beta $ and $\Psi$ is a massless scalar in higher dimensional theory . 

\item There are several families of such backgrounds with
Big-Bang/Crunch singularities, for which exponents are given in the following table \cite{Bhattacharya:2020qil}
{\tiny
\begin{table}[h]
\begin{center}
\begin{tabular}{ |c|c|c|c|c|} 
 \hline
{\bf Cosmologies} &  $k$ & $m$ & $a={\al^2\over 2}$ &  $b$  \\
 \hline 
{$AdS$ Kasner}\ & 1 & $-d_i$ & $\frac{d_i-1}{d_i}$  & $-(d_i+1)$ \\ 
 \hline
{Hv cosmology} ($z=1,\ \theta\neq 0$)\ & 1 & $-(d_i-\theta)$ & $\left(\sqrt{\frac{d_i-\theta -1}{d_i-\theta }}-\sqrt{\frac{(-\theta) }{d_i (d_i-\theta
   )}}\right)^2$ & $-\frac{(d_i-\theta)(1+d_i)}{d_i}$ \\ 
 \hline 
{Lif cosmology} ($z=d_i,\ \theta=0$)\ & 1 & $-1$ & ${d_i-1\over d_i}$ & $-3+{1\over d_i}$ \\ 
 \hline
\end{tabular}
\end{center}
\caption{Exponents for 2-dim cosmologies with ``Hv=Hyperscaling violating'' and ``Lif=Isotropic Lifshitz Kasner''.}
\label{table-exponents}
\end{table}
}
\end{itemize}

{\bf Holographic volume complexity in holographic cosmologies}
\begin{itemize}
\item  For the bulk as (\ref{dimredAnsatz}), holographic volume complexity using the complexity equals volume proposal (\ref{CV}) for the complexity surface parametrized by $t(r)$ is given by
{
\bea\label{Complexity-behavior-NHM}
& & \hskip -0.2in C = \frac{V_{d_i}}{G_{d_i+2} R} \int_\epsilon dr\,\ t(r)^{\left(k\left(\frac{(d_i+1)}{2 d_i}\right)+\frac{a}{2}\right)}\ r^{\left(m\left(\frac{(d_i+1)}{2 d_i}\right)+\frac{b}{2}\right)}\ \sqrt{{1-t'(r)^2}}\ 
\nn \\
& & \equiv\
\frac{V_{d_i}}{G_{d_i+2} R} \int_\epsilon dr\, {\cal L}\equiv {\cal L}\left(r,t(r),t'(r)\right)\,. 
\eea
} 
\item Euler Lagrange EOM for the complexity surface $t(r)$ with $k=1$ is obtained as follows
{
{
\bea\label{EOM-t[r]}
\hskip -0.2in r (a d_i+d_i+1) \left((t'(r))^2-1\right)+ (b d_i+d_i m+m)\,t(r)\,
t'(r) \left((t'(r))^2-1\right)
%\nn\\ && \hspace{40mm}   -(b d_i+d_i m+m) t'(r)
-2 d_i r\, t(r)\, t''(r) = 0. \nn \\
\eea
}
}
\vspace{-0.2in}
\item We used the following algorithm to solve (\ref{EOM-t[r]}) numerically
{
\begin{enumerate}
\item {Solve EOM for the complexity/entanglement surfaces
  semiclassically in perturbation theory using an ansatz of the form
$t(r)=t_0+\sum_{n \in \mathbb{Z}_+}c_n r^n$.}

\item  {The perturbative solutions are valid only in a certain $r$-regime,
\ie\ upto a cut-off $r_\Lambda$\ (roughly $r_\Lambda \lesssim t_0$).}
%Thus, these cannot encapsulate the entire bulk geometry.

\item {Extract initial conditions for the numerical solutions from the perturbative solutions with boundary point as $r=\epsilon=10^{-2}$. Numerical solutions are valid for the entire bulk geometry.}

%\item {\scriptsize Take boundary point as $r=\epsilon=10^{-2}$.}

\item {Numerical solutions reveal lightlike limits and the transition thereto, from
spacelike regimes near the boundary. This then allows us to numerically
evaluate holographic volume complexity/entanglement entropy and plot it against $t_0$
for various backgrounds.}

\end{enumerate}
}

\end{itemize}

{\bf Numerical results for $AdS_5$ Kasner spacetime: complexity surfaces}
\begin{itemize}
\item 
\begin{figure}[hhh]
\begin{subfigure}
  \centering
  \includegraphics[width=.45\linewidth]{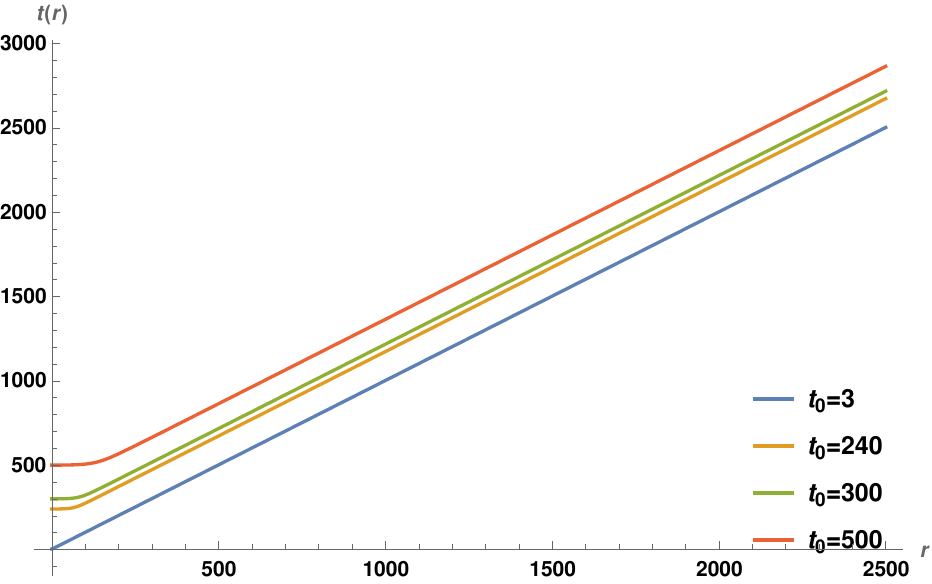}
  %\caption{1b}
  %\label{fig:sfig2}
\end{subfigure}
\begin{subfigure}
  \centering
  \includegraphics[width=.45\linewidth]{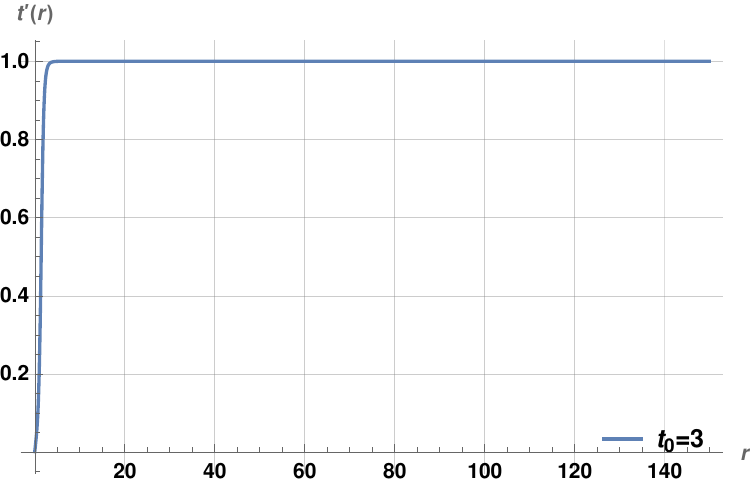}
  %\caption{1b}
  %\label{fig:sfig2}
\end{subfigure}
\begin{subfigure}
  \centering
  \includegraphics[width=.45\linewidth]{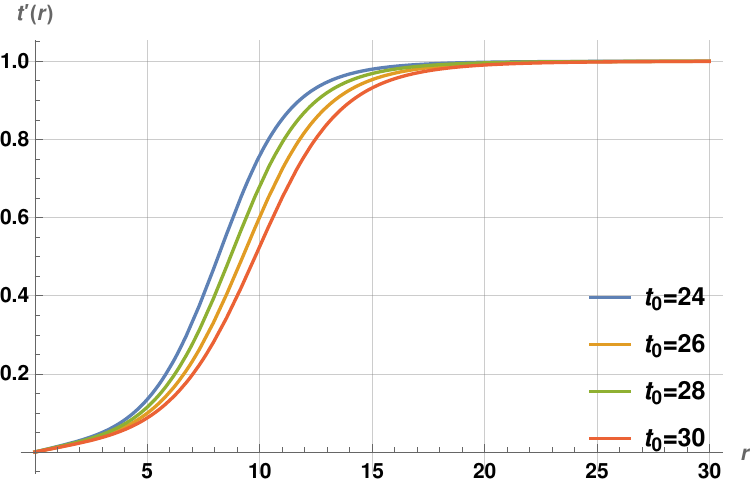}
 %\caption{1b}
  %\label{fig:sfig2}
\end{subfigure}
\begin{subfigure}
  \centering
  \includegraphics[width=.45\linewidth]{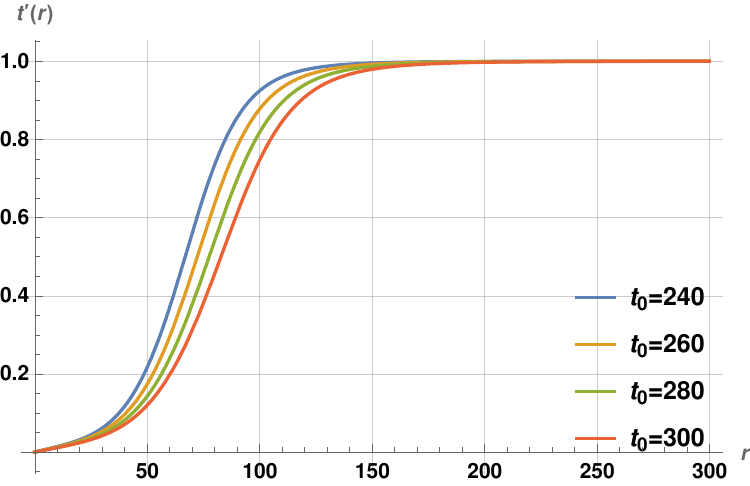}
 %\caption{1b}
  %\label{fig:sfig2}
\end{subfigure}
\caption{Numerical plots of the complexity surface versus $r$ and $t'(r)$ vs $r$ in AdS$_5$-Kasner spacetime for different slices of $t_0$. Similar results for $AdS_{4,7}$-Kasner spacetimes.}
\label{LL-i-t[r]}
\end{figure}
For $AdS_5$ Kasner spacetime, substitute $d=3$ and $a,b,m$ from Table \ref{table-exponents} in (\ref{EOM-t[r]}) which gives the EOM for complexity surfaces as: 
\begin{equation} \label{EOMtAdS5K}
r\,t(r)\, t''(r) - 4 t(r)\, t'(r)\,\left(1-t'(r)^2\right)
+ r\,\left(1 - t'(r)^2\right) = 0.
\end{equation} 
On solving this equation numerically, we obtained the plots shown in Fig. \ref{LL-i-t[r]} for $t(r)$ and its derivative $t'(r)$ as a function of radial coordinate $r$

\item From these plots in Fig. \ref{LL-i-t[r]}, we can see that complexity surfaces behave as spacelike ($t'(r)<1$) near the boundary of AdS Kasner and then they become lightlike ($t'(r)=1$) at a certain radial distance.

\end{itemize}

{\bf Holographic volume complexity of $AdS_{5}$ Kasner spacetime, numerically}
\begin{itemize}
\item Using the numerical solution of (\ref{EOMtAdS5K}), we obtained the holographic volume complexity (\ref{Complexity-behavior-NHM}) for $AdS_5$ Kasner spacetime numerically as plotted in Fig. \ref{NC-AdSK} as a function of $t_0$. We found that the dual Kasner state appears to be of vanishingly low complexity, independent of the reference state, because holographic volume complexity decreases linearly as the
anchoring time slice approaches the vicinity of the singularity,
i.e., as $t_0\rightarrow 0$.
\begin{figure}[h]
  \centering
  \includegraphics[width=.5\linewidth]{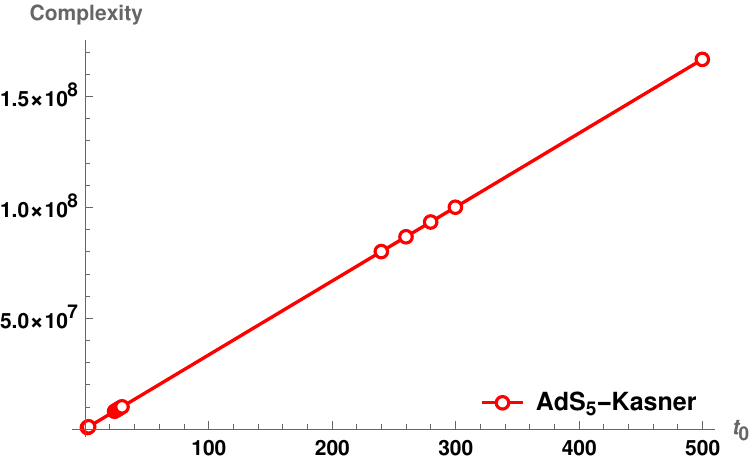}
\caption{Numerical plots of holographic volume complexity with $t_0$ in
  $AdS_{5}$-Kasner spacetimes.}
\label{NC-AdSK}
\end{figure}

\item Near boundary contribution of holographic complexity in AdS$_{d_i+2}$ Kasner with cutoff $\epsilon\equiv\Lambda_{UV}^{-1}$ is
\be\label{C-t0}
\hskip -0.5in C\ \sim\ {R^{d_i+1}\over G_{d_i+2}\,R}\, {V_{d_i}\over\epsilon^{d_i}}\,t_0\
\equiv\ N_{dof}\, V_{d_i}\Lambda_{_{UV}}^{d_i}\, t_0\,; \ \ \ \ {\rm OR} \ \ \ \ {dC\over dt_0}\ \sim\ N_{dof}\, V_{d_i}\Lambda_{_{UV}}^{d_i}\,.\nn
\ee
\end{itemize}
We also studied complexity in other cosmologies in \cite{Narayan:2024fcp}. To summarize: in AdS Kasner, hyperscaling violating and Lifshitz Kasner cosmologies:
\begin{itemize}
\item Near singularity region is the low complexity (consistent with \cite{Barbon:2015ria}\cite{Caputa:2021pad}) and less entangled state.
\begin{figure}[h] 
%\hspace{0.3pc}
\includegraphics[width=15pc]{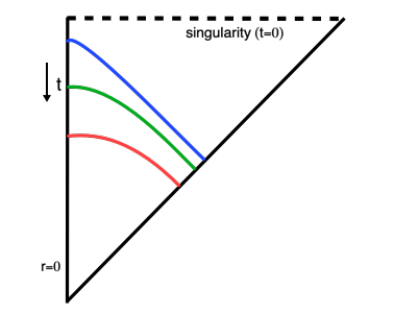}
%\hspace{0.1pc}
\begin{minipage}[b]{20.5pc}
\caption{{ \label{figbbCcmpxtySurf}
{The extremal surfaces bend
  away from the singularity (dotted line, $t=0$) and approach
  lightlike regimes eventually (approaching faster as
  $t\ra 0$).
}}}
\end{minipage}
\end{figure}

\item There are fewer degrees of freedom near the singularity region. 

\item Complexity surfaces transition from the spacelike part (near the boundary) to the lightlike part in the interior.

\item We obtained similar results for the codimension two RT/HRT entangling surfaces: {entanglement entropy}.

%\item For the finite subregion case, there is no lightlike limit for the entangling RT/HRT surfaces.

\item Spacelike singularities of this type are excluded from the entanglement wedge of the observers and hence ``{entanglement wedge cosmic censorship}''.
\end{itemize}

\subsection{Partition function and correlation functions from $AdS$ bulk}
\label{twoptAdS}
Here we will calculate the two-point function of the scalar field from the $AdS$ bulk and show that this agrees with the CFT calculation in sec. \ref{twoptcft}. {This calculation has been provided in \cite{Ammon:2015wua} explicitly}\footnote{See also \cite{Dey:2024zjx} where this calculation has been reviewed in App. {\bf B}.}. Here we will reproduce the same result from the aforementioned paper.

In Euclidean $AdS_{d+1}$, the action associated with a massive scalar field is given by
\be
\label{scactb}
S={1 \over 2G_{d+1}}\int d^{d+1}x\sqrt{g} \left(g^{\mu\nu}\partial_{\mu}\phi \partial_{\nu}\phi+M^2\phi^2\right),
\ee 
where the metric of bulk Euclidean $AdS_{d+1}$ is given below
\bea
	\label{metadsea}
	ds^2={L^2\over z^2}\left[dz^2+\sum_{i=1}^d(dx^i)^2\right].
	\eea
For convenience, it is easy to work with $G_{d+1}=1$. The equation of motion for the field $\phi$ is given as
\bea
& & \frac{1}{\sqrt{g}}\partial_\mu \left(\sqrt{g} g^{\mu \nu}\partial_\nu \phi\right)+M^2\phi=0.
\eea
In momentum space, $\phi$ has following normalised on-shell solution 
\begin{equation}
	\phi({\bf k},z) =  \frac{z^{\frac{d}{2}} K_{\nu} (kz) }{\epsilon^{\frac{d}{2}} K_{\nu} (k\epsilon)} \epsilon^{{d\over2}-\nu} \phi_b({\bf k}) \label{phisol}
\end{equation}
where $$\nu=\sqrt{{d^2\over4}+M^2 L^2}.$$
 $\epsilon$ is the cut-off value of $z$ near boundary of AdS, $\phi_b({\bf k})$ is the boundary value of scalar field and $K_{\nu} (kz)$ is modified Bessel function of second kind. Using (\ref{phisol}) and (\ref{scactb}), the on-shell action for the scalar field in momentum space is obtained as
\be
\label{onsads}
S_{\partial} = -\frac{L^{d-1}}{2} \int {d^d\mathbf{k}\over(2\pi)^d} \frac{1}{z^{d-1}} \phi({\bf k},z) \partial_z \phi(-{\bf k},z) |_{z=\epsilon}
\ee
Using the GKPW relation \cite{Gubser:1998bc}\cite{Witten:1998qj}, the partition function up to ${\cal O}(\epsilon^{2\nu})$ in $EAdS_{d+1}$ is obtained as
\be
\label{wfactads}
Z_{AdS}=\exp[-S_\del]=\exp[\frac{L^{d-1}}{2}\int \frac{d^d\mathbf{k}}{(2\pi)^d} \left\{\left(\frac{d}{2} - \nu\right) \epsilon^{-2 \nu} + k^{2\nu} \frac{a_{\nu}}{b_{\nu}} 2 \nu \right\} \phi_b({\bf k}) \phi_b (-{\bf k})]
\ee
where $a_\nu$ and $b_\nu$ are coefficients of leading modes of $K_\nu(kz)$ near boundary 
\be
\label{Kboun}
K_{\nu}(kz)=a_\nu(kz)^{\nu}+b_\nu(kz)^{-\nu}.
\ee
These are given as
\be
\label{coefads}
a_\nu=2^{-\nu -1} \Gamma[-\nu]    \hspace{50pt}   b_\nu=2^{\nu -1} \Gamma [\nu]
\ee
Using the AdS/CFT dictionary for correlation functions as discussed in sec. \ref{BBS} as
\bea
\label{nptads}
\langle O({\bf k}_1) O({\bf k}_2) \cdots O({\bf k}_n)\rangle={\delta^n \log Z\over \delta {\phi}_b({\bf k}_1) \delta{\phi}_b({\bf k}_2) \cdots \delta {\phi}_b({\bf k}_n)}.
\eea
Two point correlation function in momentum space is obtained as
\begin{equation}
	\label{adstwopt}
	\langle O(\mathbf{k}) O(-\mathbf{k}) \rangle_{AdS} = -  \frac{\pi}{2^{2\nu-1} \Gamma[\nu]^2 \sin(\pi \nu)} k^{2\nu} L^{d-1}.
\end{equation}
Similarly, one can obtain the partition function when $\nu$ is an integer as\footnote{{When you Fourier transform $k^2$, it becomes a box operator. This means that when $\nu$ is an integer and you are doing a Fourier transform, all you end up getting is powers of the box acting on the delta function. So you need the log term to give you a non-local answer.}}
\begin{multline}
	\label{partadsp2}
	\log Z_{AdS}=\frac{L^{d-1}}{2}\int \frac{d^d\mathbf{k}}{(2\pi)^d} \left[\left(\frac{d}{2}-\nu\right)\epsilon^{-2\nu}+\left\{\frac{\tilde{b}_0}{\tilde{a}_0}(1+2\nu \log(\epsilon))+2\nu\frac{\tilde{c}_0}{\tilde{a}_0}\right\}k^{2\nu}\right.\\
	\left.+2\nu\frac{\tilde{b}_{0}}{\tilde{a}_{0}} k^{2\nu}\log(k) \right] \phi_b({\bf k}) \phi_b (-{\bf k})
\end{multline}
where 
\begin{align}
	\tilde{a}_0=2^{\nu-1}\Gamma[\nu]; \hspace{20pt}
	\tilde{b}_0=\frac{(-1)^{\nu-1}}{2^{\nu}\Gamma[\nu+1]};\hspace{20pt}
	\tilde{c}_0=\frac{(-1)^{\nu+1}}{2^{\nu}\Gamma[\nu+1]}\left(\gamma_E-\frac{1}{2}\sum_{m=1}^{d/2}\frac{1}{m}-\log(2)\right).
\end{align}
In the above equation, $\gamma_E$ is the Euler number. Two point correlation function in momentum space for integer $\nu$ is given by
\begin{align}
	\langle O(\mathbf{k})O(\mathbf{-k})\rangle_{AdS}=-\frac{(-1)^{\nu} }{2^{2\nu-2}\Gamma[\nu]^2}L^{d-1}k^{2\nu}\log(k).\label{okokint}
\end{align}	
Equations (\ref{adstwopt}) and (\ref{okokint}) takes the following form in position space
\begin{equation}
	\label{oxoyads}
	\langle O(\mathbf{x})O(\mathbf{y})\rangle_{AdS}= \frac{2\nu }{\pi^{d\over2}}\frac{\Gamma[\frac{d}{2}+\nu]}{\Gamma[\nu]}{L^{d-1}\over|{\bf x}-{\bf y}|^{d+2\nu}}
\end{equation}
which is same as two point function of CFT in $d$ dimensions obtained in (\ref{twoptcftf}) with the following identifications
\bea
& & 2\Delta_1=d+2\nu \, , \nn \\
& & d_{12}=\frac{2\nu }{\pi^{d\over2}}\frac{\Gamma[\frac{d}{2}+\nu]}{\Gamma[\nu]}L^{d-1} \, .
\eea
Hence, we found the same two-point correlators from CFT as well as bulk AdS calculation. In this lecture, we have discussed only three consistency checks. There are many more, see for example a list partial references \cite{NCM, Blake:2022uyo, Doucot:2024hzq, Yadav:2023glu, Bhattacharyya:2007vjd} apart from other references that we mentioned earlier.

\section{Generalization of the AdS/CFT correspondence}

\subsection{Holography in de Sitter space}
\label{dSholography}
Holography in AdS spacetime has been explored extensively. Our universe behaves like a de Sitter space, so one can ask the question of how holography works in de Sitter space. There are various proposals, such as the dS/CFT correspondence \cite{Strominger:2001pn}\cite{Maldacena:2002vr}, static patch holography \cite{Susskind:2021omt}, and the DS/dS correspondence \cite{Alishahiha:2004md}\cite{Alishahiha:2005dj}. We will discuss briefly these proposals. First, let us understand the de Sitter space. We are following \cite{Erfani:2020wev} to discuss this.

{\bf de Sitter space}: de Sitter space is the solution of Einstein's EOM with positive cosmological constant ($\Lambda >0$). We can embedd $d$ dimensional de Sitter space in $(d+1)$ dimensional Minkowski spacetime using the following embedding equation
\bea
& & -X_0^2+\sum_{i=1}^d X_i^2=l^2.
\eea
The metric of Minkowski spacetime in $(d+1)$ dimensions is given by
\bea \label{mdSM}
& & ds^2=-dX_0^2+\sum_{i=1}^d dX_i^2.
\eea
Below, we will derive the metric for de Sitter space in four dimensions. For higher dimensions, the discussion will follow up on this, similar to AdS spacetime in sec. \ref{AdSmd}.\\

\begin{figure}[h]
\begin{center}
\includegraphics[width=0.45\textwidth]{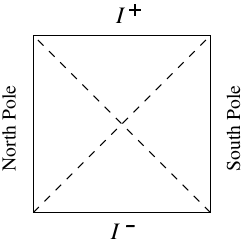}
\end{center}
\caption{Penrose diagram of de Sitter space. This figure is taken from \cite{Spradlin:2001pw}.}
\label{pddS}
\end{figure}

{\bf Static patch}: The Static patch metric of four de Sitter space can be derived using the following embedding coordinates
\bea \label{XisdSSP}
& & X_0= l\sqrt{1-\frac{r^2}{l^2}} \sinh\left(\frac{t}{l}\right)\, , \nn \\
& & X_1= l\sqrt{1-\frac{r^2}{l^2}} \cosh\left(\frac{t}{l}\right) \, , \nn \\
& & X_2= r \sin \theta \cos\phi \, , \nn \\
& & X_3= r \sin \theta \sin\phi \, , \nn \\
& & X_4= r \cos\theta \,  . 
\eea
Substituting $X_i$s from (\ref{XisdSSP}) into (\ref{mdSM}), one obtain the metric of static patch of $dS_4$ as 
\bea
& & ds^2=-\left(1-\frac{r^2}{l^2}\right)dt^2+\frac{dr^2}{\left(1-\frac{r^2}{l^2}\right)}+r^2 d\Omega_2^2 \, ,
\eea 
where $d\Omega_2^2=d\theta^2+\sin^2\theta d\phi^2$.\\
~\\
{\bf Global de Sitter}: The embedding coordinates for global $dS_4$ are given as follows
\bea \label{XisgdS}
& & X_0= l \sinh\left(\frac{\tau}{l}\right)\, , \nn \\
& & X_1= l \cosh\left(\frac{\tau}{l}\right) \cos\theta \, , \nn \\
& & X_2= l \cosh\left(\frac{\tau}{l}\right) \sin\theta \cos\phi \, , \nn \\
& & X_3= l \cosh\left(\frac{\tau}{l}\right) \sin\theta \sin\phi \cos\alpha \, , \nn \\
& & X_4= l \cosh\left(\frac{\tau}{l}\right) \sin\theta \sin\phi \sin\alpha \,  . 
\eea
Upon substution of $X_i$s from (\ref{XisgdS}) in (\ref{mdSM}), we obtain the metric of global $dS_4$ spacetime as below
\bea \label{gdS4}
& &  ds^2=-d\tau^2+l^2 \, \cosh^2\left(\frac{\tau}{l}\right)d\Omega_3^2,
\eea
where $d\Omega_3^2$ is the metric of $S^3$ and is given as: $d\Omega_3^2=d\theta^2 +\sin^2\theta\left(d\phi^2+\sin^2\phi d\alpha^2\right)$.
For the global $dS_d$ spacetime, embedding coordinates are
\bea \label{Xisd}
& & X_0= l \sinh\left(\frac{\tau}{l}\right) \, , \nn \\
& & X_j=l \cosh\left(\frac{\tau}{l}\right) \omega_j \, , \nn \\
\eea
where $\omega_j$ are the coordinates of $S^{d-1}$. Substituting (\ref{Xisd}) in (\ref{mdSM}), we obtain the metric of global de Sitter spacetime in $d$ dimensions as
\bea \label{gdSd}
& &  ds^2=-d\tau^2+l^2 \, \cosh^2\left(\frac{\tau}{l}\right)d\Omega_{d-1}^2 =-d\tau^2+a(\tau)^2\Omega_{d-1}^2 ,
\eea
where $a(\tau)=l \cosh\left(\frac{\tau}{l}\right)=\frac{l}{2} \left(e^{\frac{\tau}{l}}+e^{-\frac{\tau}{l}} \right)$ is the scale factor. 
\begin{figure}[h]
\begin{center}
\includegraphics[width=0.5\textwidth]{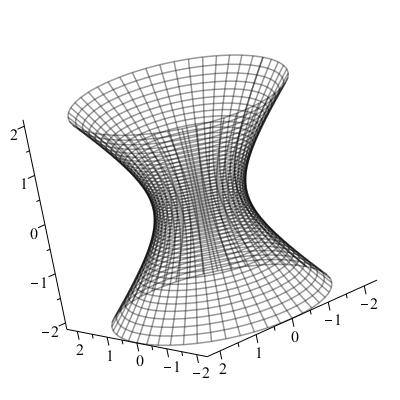}
\end{center}
\caption{Global dS spacetime. This figure is taken from \cite{Erfani:2020wev}.}
\label{gdSfig}
\end{figure}
Let us analyze the behavior of the scale factor
\bea
& & a(\tau \rightarrow 0)= l \, \nn \\
& & a(\tau \rightarrow \infty )= \frac{l}{2} e^{\infty} \, \nn \\
& &  a(\tau \rightarrow -\infty)= \frac{l}{2} e^{\infty} \, . \nn \\
\eea
Therefore, the global de Sitter space can be viewed as a sphere of large size in the past infinity ($\tau \rightarrow -\infty$) and then becomes a sphere of minimum size at $\tau = 0$ and then again becomes a sphere of large size in the future infinity ($\tau \rightarrow \infty$). See Fig. \ref{gdSfig} for the pictorial representation. Penrose diagram of de Sitter space is shown in Fig. \ref{pddS}.
In the penrose diagram, $I^+$ ($r \rightarrow \infty$) and $I^-$ ($r \rightarrow \infty$) are future and past boundaries. North Pole and South Pole correspond to $\theta=0$ (or $r=0$ in static coordinates) and $\theta=\pi$ (or $r=0$ in static coordinates), respectively. The dotted lines are cosmological horizons $r=l$. The region $0 \leq r \leq l$ is the static patch of de Sitter space.

 We will start with dS/CFT correspondence \cite{Strominger:2001pn}\cite{Maldacena:2002vr}, for which, first, we need to understand the Hartle-Hawking proposal to compute the wavefunction of the universe \cite{HH}.

\begin{figure}[h]
\begin{center}
\includegraphics[width=0.3\textwidth]{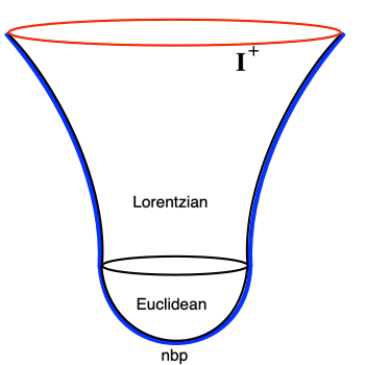}
\end{center}
\caption{No-boundary de Sitter space, with the top Lorentzian region continuing smoothly into the Euclidean hemisphere region ending at the no-boundary
point. The blue surface is the no-boundary extremal surface for the full subregion. The picture is taken from \cite{Goswami:2024vfl}.}
\label{dSnb}
\end{figure}

{ The Hartle-Hawking proposal or the no boundary proposal is a method for determining the wavefunction of the universe, treating it as a function of the three-dimensional spatial geometry ($h_{ij}$) and the values of various fields ($\phi$) defined on that geometry. This approach involves a generally complex four-dimensional geometry that terminates at the specified three-dimensional spatial surface without any additional boundaries \cite{Halliwell:2018ejl}. Notably, it lacks a boundary into the past as well. Based on this no-boundary geometry, the wavefunction is then computed via}
\bea \label{WF}
& & \Psi[h_{ij},\phi] \propto e^{I[g_{\mu \nu},\Phi]},
\eea
where $I$ represents the classical action computed for the no-boundary four-dimensional geometry, defined by the metric $g_{\mu \nu}$ and fields $\Phi$. No boundary geometry in de Sitter space is obtained as follows. Consider the de Sitter space, cut in the middle, and glue this with the bottom hemisphere; see Fig \ref{dSnb}.
%$g_{\mu \nu}$​ %and the fields $\phi$.}
Relating (\ref{WF}) with Fig. \ref{dSnb}: $g_{\mu \nu}$ and $\Phi$ are the metric of the de Sitter space and scalar field in four-dimensional no-boundary de Sitter space, whereas $h_{ij},\phi$ are defined on the three-sphere existing at the future boundary ($I^+$) of de Sitter space and wavefunction of the universe is denoted by $\Psi_{dS}$. 

{\bf dS/CFT correspondence}: In dS/CFT duality \cite{Strominger:2001pn}\cite{Maldacena:2002vr}, dual field theory lives at the future boundary of de Sitter space, which has imaginary central charge, i.e., the dual theory is nonunitary. The precise dS/CFT dictionary is $Z_{CFT}=\Psi_{dS}$ where $Z_{CFT}$ is the partition of Euclidean CFT at the future boundary and $\Psi_{dS}$ is the no boundary Hartle-Hawking state \cite{HH}. Therefore, ``gravity living in no boundary de Sitter space is dual to CFT living at the future boundary of de Sitter space''. The analytical solution for the slow-roll correction to the de Sitter space has been obtained in \cite{Maldacena:2024uhs}, which is useful to study cosmology from an analytical approach: earlier, we had a numerical approach. For the quantum information aspects of slow-roll no-boundary de Sitter space, see \cite{Goswami:2024vfl}. In these cases, dual theory is not well understood.\\
~\\
{\bf Excercise}: Compute the pseudo entropy in $dS_3/CFT_2$ correspondence by following \cite{Narayan:2023zen}. \\
~\\

{\bf Static patch holography}: In static patch holography, one natural question was asked: where should we locate the holographic screen so that the maximum entropy of the spatial region described by the hologram is sufficient to encode everything in the geometry? The answer to this question comes from Bousso wedges, which are causal light sheet regions associated with a surface along non-expanding null directions. This defines where we can locate the holographic screen. If we consider a holographic screen near the poles, then we will not be able to encode everything inside the static patch of de Sitter space. Therefore, we need to consider a holographic screen close to the cosmic horizon. See \cite{Susskind:2021omt} for more discussion on this.

\begin{figure}[h]
\begin{center}
\includegraphics[width=0.7\textwidth]{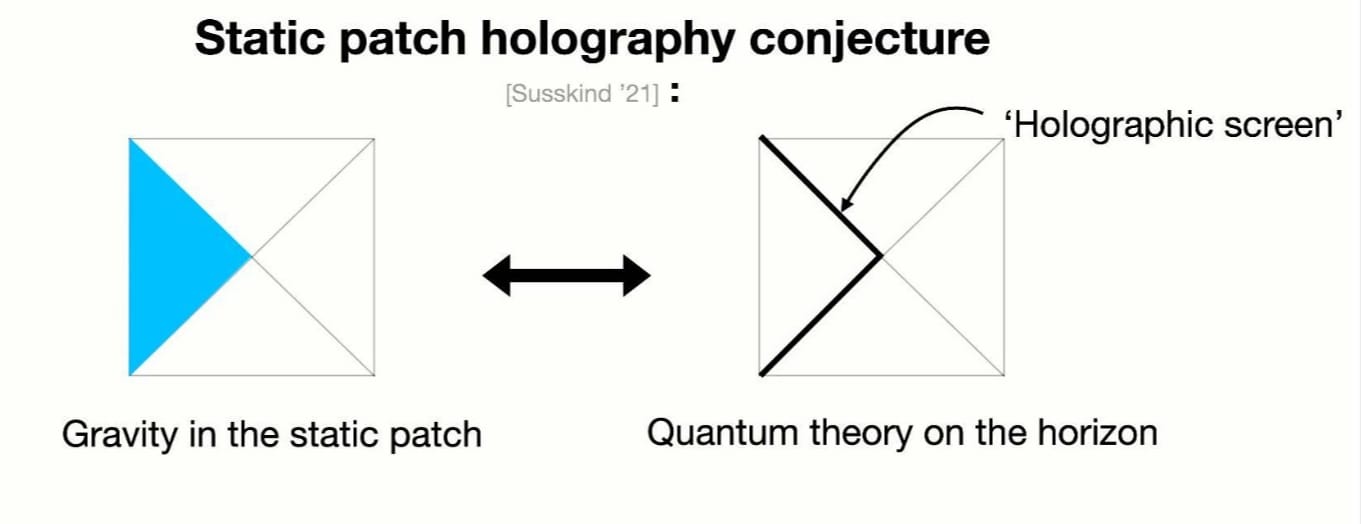}
\end{center}
\caption{Illustration of static patch holography. This picture is taken from  \url{https://pdf.pirsa.org/files/24100113.pdf}.}
\label{SPH}
\end{figure}

In static patch holography \cite{Susskind:2021omt} (see also \cite{Franken:2023pni}), the hologram is considered as a stretched horizon (surface near the cosmolgical horizon stretching between future and past boundaries of de Sitter space) where boundary theory lives, and it is assumed that the static patch of de Sitter space is dual to boundary theory living at this stretched horizon. See Fig. \ref{SPH} for pictorial description.

{\bf DS/dS correspondence}: The DS/dS correspondence \cite{Alishahiha:2004md}\cite{Alishahiha:2005dj} states that ``{\it static patch of de Sitter space in $d$ dimensions is dual to two $d-1$ dimensional IR CFTs which are coupled to each other and $d-1$ dimensional de Sitter gravity living at the central slice $r=\frac{\pi l}{2}$}.'' This can be understood as follows: AdS and dS spaces can be foliated with dS slices
\bea
& & ds^2_{(A)dS_d}=dr^2+\sin(h)^2\left(\frac{r}{l}\right)ds^2_{dS_{d-1}},
\eea
where $r$ is the radial coordinated and $l$ is curvature length. For $AdS_d$, $r\in \left(-\infty,\infty\right)$ and for $dS_{d}$, $r\in \left[0,\pi l\right]$. The UV conformal boundary of $AdS_d$ is located at $r=\pm \infty$ and IR at $r=0$. The near-horizon region of $DS_d$ at $r=0,\pi l$ is isomorphic to $AdS_d$ foliated with $dS_{d-1}$ slices in the IR, and hence we have two CFTs on $dS_{d-1}$ in the IR in DS/dS correspondence. The central charge is real in the $DS/dS$ correspondence. 

So far, we have discussed various versions of codimension one holography in de Sitter space. For codimension two holography in de Sitter space, see \cite{Yadav:2024ray}, where one can see that the aforementioned three proposals of holography in de Sitter space can appear in a single framework, which we term as ``DS wedge holography''.

\subsection{Review: double holography and wedge holography}
Since in this lecture notes, we would like to discuss flat space holography from the wedge holography perspective. Therefore first we review what is wedge holography in AdS spacetime and then we proceed to flat spacetime in the next section.\\
~\\
{\bf AdS/BCFT correspondence}: The holographic dual of a boundary conformal field theory (BCFT)\footnote{A conformal field theory defined on a manifold $M$ with boundary $\partial M$.}
is given by anti de Sitter (AdS) spacetime with an end of the world (EOW) brane $Q$ \cite{Takayanagi:2011zk, Fujita:2011fp}. The presence of a boundary in a $d$-dimensional conformal field theory explicitly
breaks the global conformal symmetry group from $SO(2,d)$ to $SO(2,d-1)$.
We now explain this symmetry breaking and its holographic realization.

A $d$-dimensional conformal field theory (CFT) without boundaries possesses the
global conformal symmetry group $SO(2,d)$ generated by translations $P_\mu$, Lorentz transformations $M_{\mu\nu}$, dilatations $D$, and special conformal transformations $K_\mu$. We introduce a planar boundary located at
\begin{equation}
x_\perp = 0,
\end{equation}
where $x_\perp$ denotes the coordinate normal to the boundary, and
$x^i_{\parallel}$ ($i=1,\dots,d-1$) denote the coordinates parallel to the boundary.
Only those conformal transformations that preserve the boundary
$x_\perp=0$ remain as symmetries of the theory. The preserved generators are
\begin{equation}
\{ P_i,\; M_{ij},\; D,\; K_i \}, \qquad i,j = 1,\dots,d-1,
\end{equation}
corresponding to translations, rotations, dilatations, and special conformal
transformations parallel to the boundary.
The generators broken by the presence of the boundary are
\begin{equation}
P_\perp, \qquad M_{i\perp}, \qquad K_\perp .
\end{equation}

The unbroken generators close into the conformal algebra in $(d-1)$ dimensions,
\begin{equation}
SO(2,d) \;\longrightarrow\; SO(2,d-1),
\end{equation}
so that the global symmetry group of a boundary conformal field theory (BCFT) is
$SO(2,d-1)$. This residual symmetry acts as the conformal group on the
$(d-1)$-dimensional boundary, while bulk operators transform in representations of
this reduced conformal algebra.
\begin{figure}[h]
\begin{center}
\includegraphics[width=0.5\textwidth]{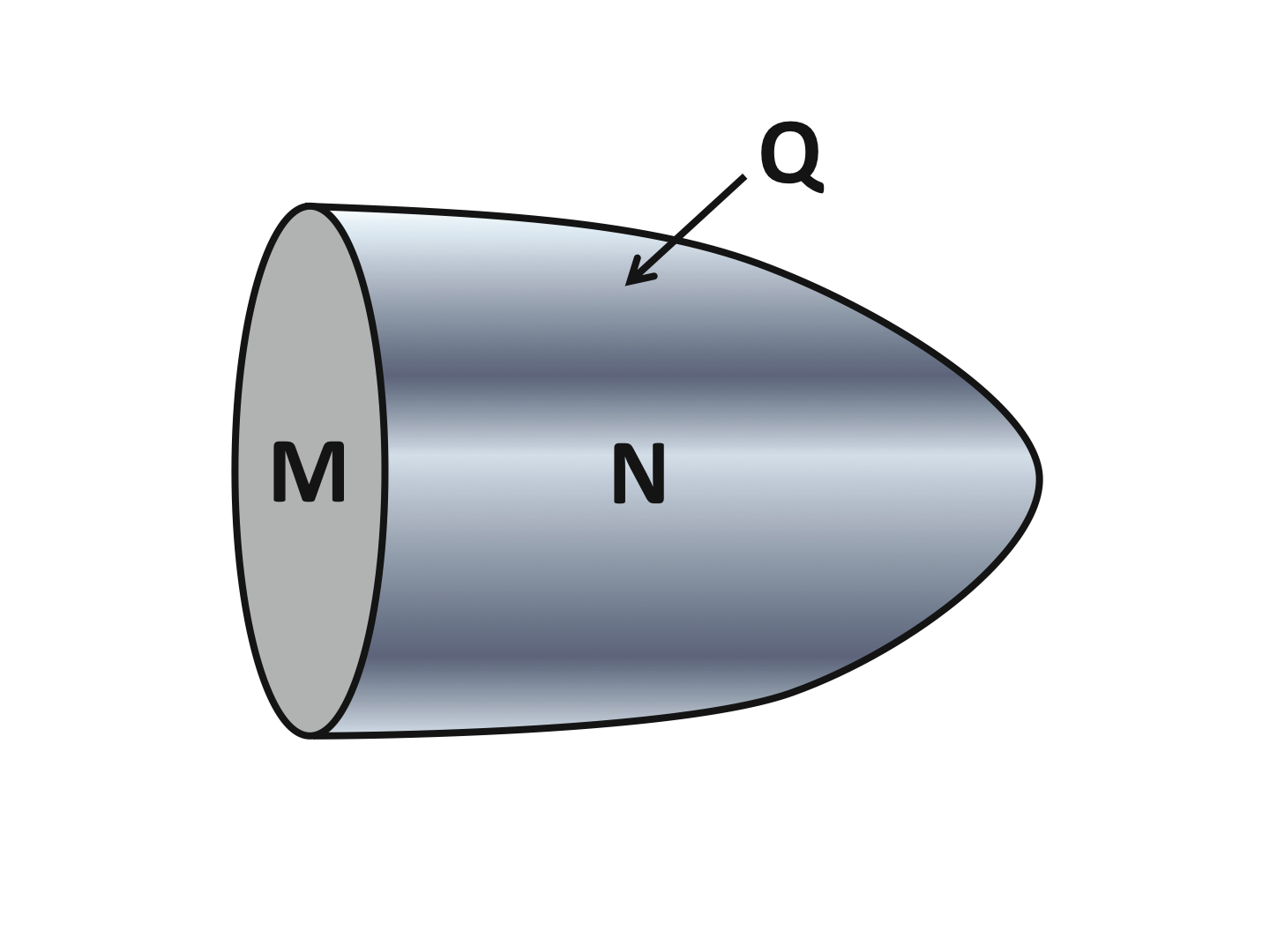}
\end{center}
\caption{Illustration of $AdS/BCFT$ correspondence. This figure is taken from \cite{Fujita:2011fp}.}
\label{ABC}
\end{figure}
To construct the gravity dual of a $d$ dimensional BCFT, we extend the $d$ dimensional manifold $M_d$ into a $(d+1)$ dimensional bulk spacetime $N_{d+1}$ such that the boundary of the bulk geometry satisfies
\begin{equation}
\partial N_{d+1} = M \cup Q ,
\end{equation}
where $M$ corresponds to the asymptotic AdS boundary and $Q$ is the EOW brane. See Fig. \ref{ABC} for the pictorial illustration. 

We consider the gravitational action
\bea \label{IBCFT}
& & I=\frac{1}{16 \pi G_N}\int_N d^{d+1}x \sqrt{-g}\,\left(R-2\Lambda\right)
+\frac{1}{8 \pi G_N} \int_Q d^d x \sqrt{-h} \, \left(K-T\right),
\eea
where $R$, $\Lambda$, and $G_N$ denote the Ricci scalar, cosmological constant,
and Newton's constant, respectively.
Further, $h$, $K$, and $T$ are the induced metric, trace of the extrinsic curvature,
and the tension of the EOW brane.

Varying the action \eqref{IBCFT} with respect to the induced metric on the EOW brane
yields
\bea
& & \delta I=\frac{1}{16 \pi G_N}\int_Q d^d x \sqrt{-h}
\left( K_{ab}- (K-T) \, h_{ab}\right)\delta h^{ab}.
\eea

If we impose Dirichlet boundary conditions on $Q$, then $\delta h^{ab}=0$.
For Neumann boundary conditions (NBC), we obtain
\bea
& & K_{ab}- (K-T) \, h_{ab}=0.
\eea

Taking the trace of this equation gives the tension of the EOW brane as follows
\bea \label{NBC}
& & h^{ab}K_{ab}- (K-T) \,h^{ab} h_{ab}=0 \nn \\
& & \implies K-(K-T) \,d=0 \nn \\
& & \implies T=\left(\frac{d-1}{d}\right) K.
\eea

The holographic dual of BCFT is obtained as follows.
Since the bulk theory is Einstein gravity with a negative cosmological constant,
we localize gravity on the EOW brane, a setup known as braneworld holography.
The gravity living on the EOW brane is further dualized using the AdS/CFT
correspondence. Since holography is applied twice, this construction is known as
\emph{double holography}. We consider the bulk metric in the following form
\bea \label{BMDH}
& & ds^2=d\rho^2+\cosh^2\left(\frac{\rho}{L}\right)ds^2_{AdS_d},
\eea
where $\rho \in (-\infty,\infty)$ and constant $\rho$ slices correspond to
$d$ dimensional AdS spacetime. The $d$ dimensional AdS metric is given by
\bea
& & ds^2_{AdS_d}=\frac{L^2}{y^2}\left[-dt^2+dy^2+d\vec{\omega}^2\right],
\eea
with $\vec{\omega} \in \mathbb{R}^{d-2}$. In $(d+1)$ dimensions, the cosmological constant is
\bea
& & \Lambda=-\frac{d(d-1)}{2 L^2}.
\eea
Introducing new coordinates $z$ and $x$ as below
\bea
& & z=\frac{y}{\cosh\left(\frac{\rho}{L}\right)}, \qquad
x=y \tanh\left(\frac{\rho}{L}\right)\,.
\eea
Using the above, the metric (\ref{BMDH}) becomes
\bea
& & ds^2=\frac{L^2}{z^2}\left[dz^2-dt^2+dx^2+d\vec{\omega}^2\right].
\eea

To construct the gravity dual of BCFT, we place the EOW brane $Q$ at $\rho=\rho_*$
and restrict the range of $\rho$ to $-\infty < \rho < \rho_*$.
For the metric \eqref{BMDH}, the extrinsic curvature is
\bea
& & K_{ab}=\frac{1}{L}\tanh\left(\frac{\rho}{L}\right).
\eea

Using this expression together with the Neumann boundary condition
\eqref{NBC}, we obtain the tension of the EOW brane as
\bea
& & T=\left(\frac{d-1}{L}\right)\tanh\left(\frac{\rho_*}{L}\right).
\eea
~\\
{\bf Wedge holography}:
\begin{figure}[h]
\begin{center}
\includegraphics[width=0.7\textwidth]{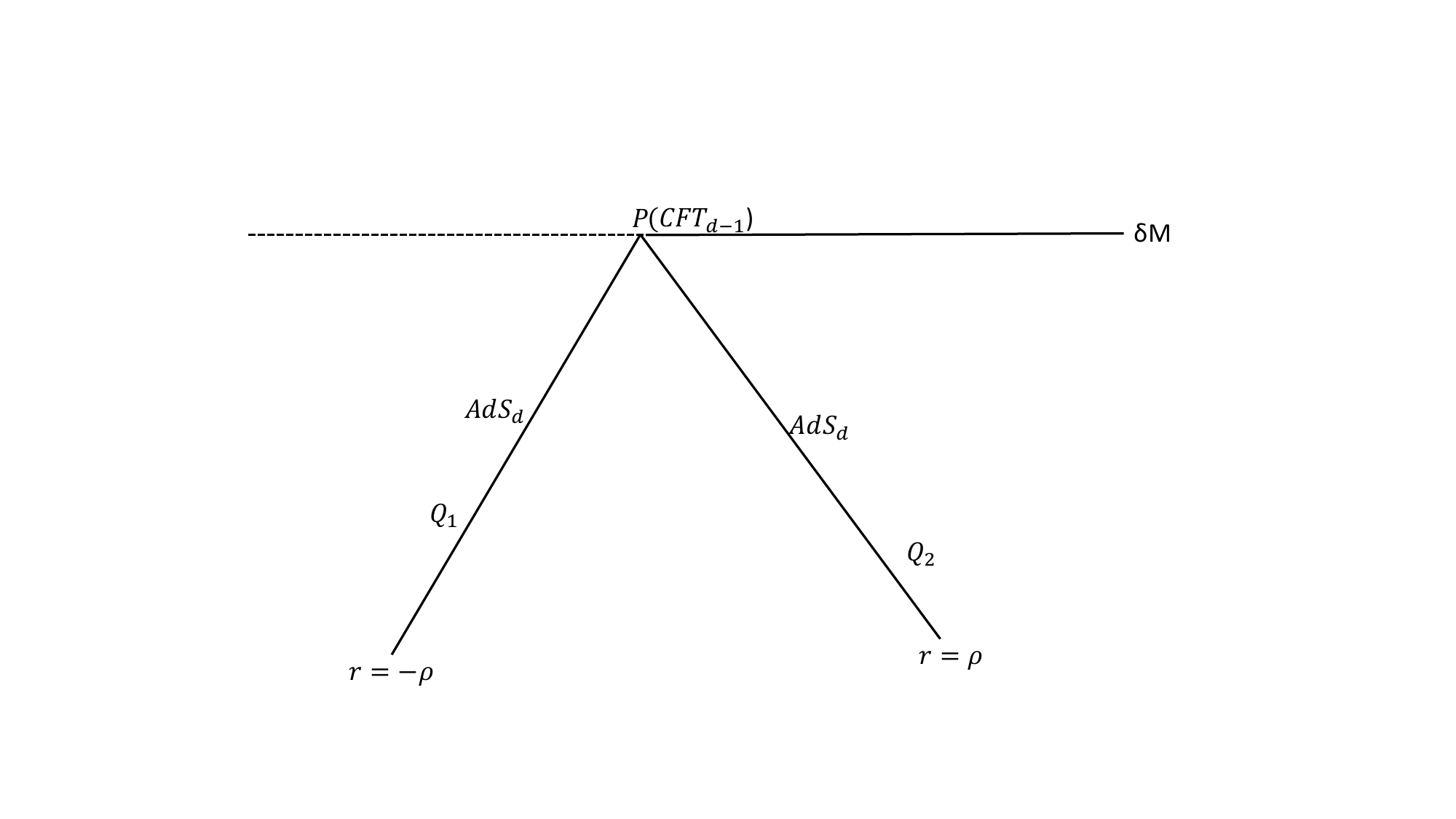}
\end{center}
\caption{Illustration of wedge holography. $r$ is the radial direction which provide location of the branes ($r=\pm \rho$) which can be seen from (\ref{BMDH}) by replacing $\rho$ with $r$ and boundary term of (\ref{IBCFT}) will contain two terms for two branes. This figure is taken from \cite{Yadav:2023qfg}.}
\label{WH}
\end{figure}
 Wedge holography\footnote{For AdS spacetime, the construction of wedge holography was first developed in \cite{Akal:2020wfl, Miao:2020oey}. This framework was subsequently extended to flat spacetime in \cite{Ogawa:2022fhy}, where holographic dualities were explored in the absence of a cosmological constant. More recently, wedge holography has been generalized to de Sitter (dS) spacetime in \cite{Yadav:2024ray}, providing a codimension two holographic description of spacetime with positive cosmological constant.} provides an example of a co dimension two holographic correspondence. Realization of wedge holography can be understood in a two-steps: (i) one considers the localization of $(d+1)$-dimensional Einstein gravity onto $d$-dimensional Karch-Randall branes embedded in the bulk. This mechanism, known as braneworld holography \cite{Karch:2000ct, Karch:2000gx}, yields an effective gravitational theory on each brane, (ii) since the induced geometry on the Karch-Randall branes is asymptotically AdS, the gravitational dynamics on the branes admit a dual description in terms of a conformal field theory (CFT) living on their $(d-1)$-dimensional boundaries, in accordance with the AdS/CFT correspondence \cite{Maldacena:1997re}.

Combining these two steps, one obtains wedge holography. The wedge holographic dictionary is as follows: ``$(d+1)$-dimensional Einstein gravity in the bulk region bounded by the Karch-Randall branes (wedge region) is dual to a $(d-1)$-dimensional defect CFT living at the intersection (corner) of the branes''. See Fig. \ref{WH} for the pictorial representation.

\subsection{Flat space holography from wedge holography perspective}
\label{fsholography}
In this section, we aim to discuss holography for flat spacetime ($\Lambda=0$). There are various progresses in this direction, see for example \cite{Ball:2019atb,Cheung:2016iub,Nguyen:2022zgs,Donnay:2022hkf,Laddha:2022nmj}. In this lecture, we will focus on just one case, which is wedge holography in flat spacetime, and it is based on the paper \cite{Ogawa:2022fhy}\footnote{In this section, figures are also taken from \cite{Ogawa:2022fhy}.}.

\subsubsection{Hyperbolic and de Sitter slicing of flat spacetime}
\label{sec:setup}
Let us begin with a flat spacetime of dimension $d+1$, namely $\mathbb{R}^{1,d}$:
\be \label{FSM}
ds^2=-dT^2+dR^2+R^2d\Omega_{d-1}^2.
\ee
The spacetime admits a decomposition into hyperbolic slices $H^d$ and de Sitter slices $dS^d$, a structure that points toward holography \cite{deBoer:2003vf}.

The hyperbolic slicing is obtained by the coordinate transformation
\ba
T=\eta\cosh\rho, \  \ R=\eta\sinh\rho \, ,
\ea
which brings the metric (\ref{FSM}) into the form 
\ba
&& ds^2=-d\eta^2+\eta^2(d\rho^2+\sinh^2\rho d\Omega_{d-1}^2),\ \ \  \mbox{[hyperbolic patch]}, 
\label{hypatch}
\ea

Alternatively, the de Sitter slicing arises from the parametrization
\ba
T=r\sinh t,\ \ \ R=r\cosh t,
\ea
leading to the metric (\ref{FSM}) as
\ba
&& ds^2=dr^2+r^2(-dt^2+\cosh^2 t d\Omega_{d-1}^2). \ \ \ \  \mbox{[de Sitter patch]}, 
\label{dspatch}
\ea
\begin{figure}[h]
  \centering
  \includegraphics[width=7cm]{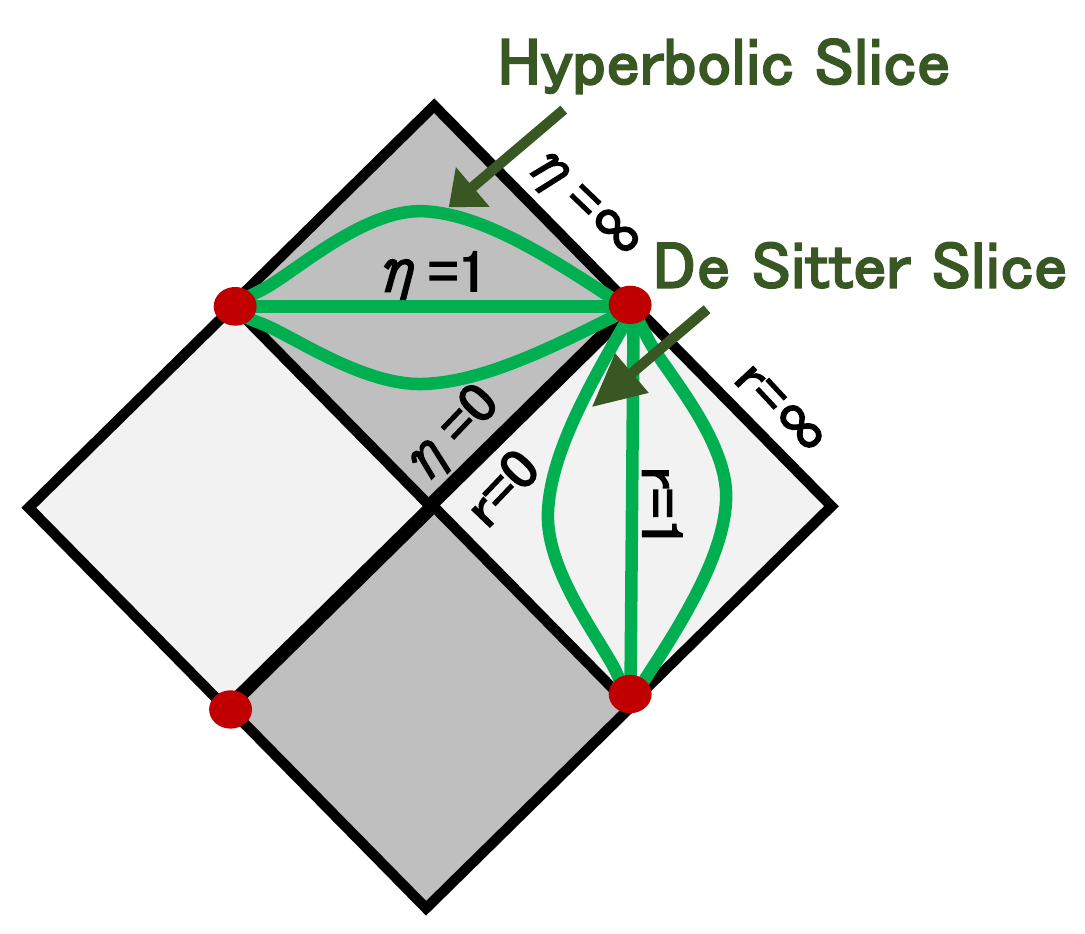}
  \includegraphics[width=5.8cm]{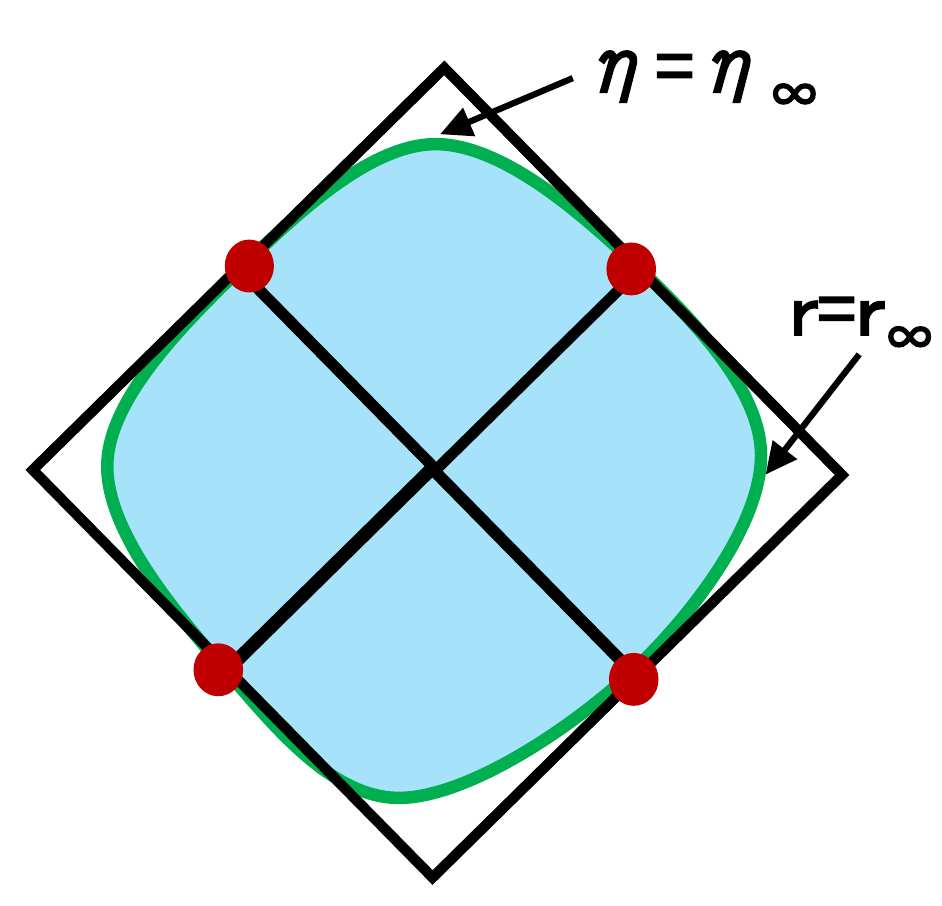}
  \caption{Hyperbolic and de Sitter slices in Minkowski Space (left) and its regularization (right).}
\label{fig:setupp}
\end{figure}
In both coordinate systems the radial variables ($\eta$ and $r$) range as 
$0\leq \eta<\infty$  and $0\leq r<\infty$. By gluing the two patches smoothly at 
$\eta=0$ and $r=0$, one recovers the complete four-dimensional Minkowski spacetime, as illustrated in the left panel of Fig.~\ref{fig:setupp}.

We impose a regularization on the coordinates $\eta$ and $r$ as
\ba
0\leq \eta\leq\eta_\infty,\ \ \ \ \ \ 0\leq r\leq r_\infty. \label{wedger}
\ea
so that the hyperbolic and de Sitter patches are effectively compactified to $H^d$ and $dS_d$, in direct analogy with wedge holography in AdS \cite{Akal:2020wfl}\cite{Miao:2020oey}, which itself is viewed as a doubled extension of AdS/BCFT \cite{Takayanagi:2011zk,Fujita:2011fp}. Extending this wedge holographic construction to ($d+1$)-dimensional Minkowski spacetimes suggests a possible ``{\it duality between a $(d-1)$-dimensional CFT living on $S^{d-1}$ and gravity in the wedge region (\ref{wedger})}''. As in the AdS/CFT correspondence \cite{Maldacena:1997re} and dS/CFT duality \cite{Strominger:2001pn}, it is natural to introduce a UV cutoff in the boundary CFT, corresponding to the geometric cutoff 
\ba
\rho\leq\rho_\infty,\ \ \  t\leq t_\infty.  \label{cutoffb}
\ea

In what follows, we analyze the hyperbolic and de Sitter slices independently via the application of wedge holography.

\subsubsection{Wedge holography for hyperbolic slices}\label{sec:HP}
%%%%%%%%%%%%%%%%%%%%%%%%%%%%%%%%%%%%%%%%%%%%%%%%%

\begin{figure}[hhh]
  \centering
  \includegraphics[width=12cm]{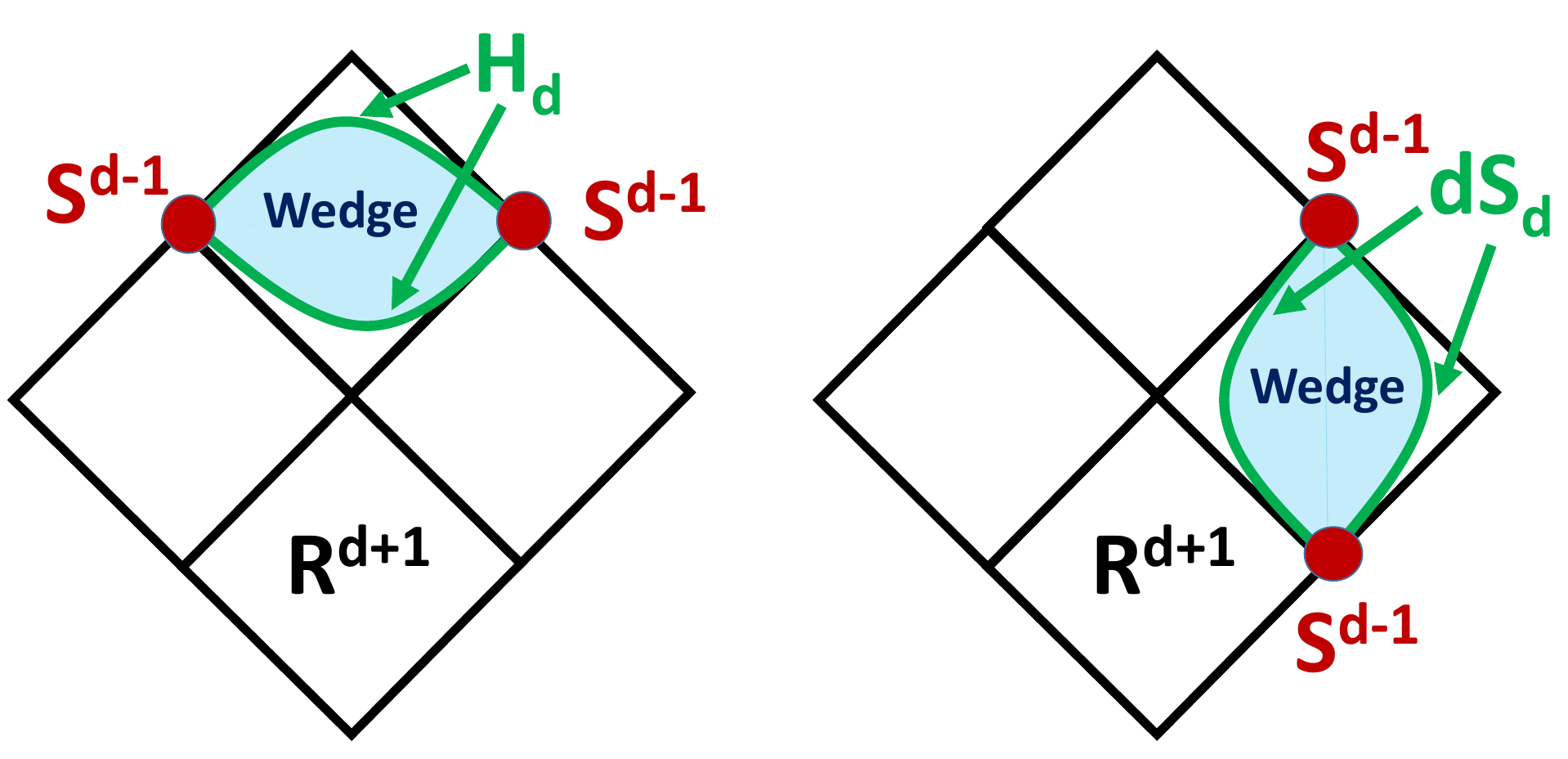}
  \caption{Sketches of two types of codimension two holographic dualities in flat space. The whole diamond describes a $ d+1$-dimensional flat spacetime. The left and right panels describe the hyperbolic and de Sitter sliced wedges (blue regions) surrounded by two end-of-the-world branes (green surfaces), respectively. The authors argued that each of them is dual to a CFT on the $d-1$ dimensional sphere (red points).} 
\label{fig:wedgeah}
\end{figure}
The authors began by formulating wedge holography for the hyperbolic slicing, illustrated in the left panel of Fig.~\ref{fig:wedgeah}. The $(d+1)$-dimensional wedge $W^h$ is defined by restricting the coordinate $\eta$ in the hyperbolic patch (\ref{hypatch}) to the range
\ba
\eta_1\leq \eta\leq \eta_2. \label{etawdge}
\ea
At the two boundaries $\eta=\eta_1$ and $\eta=\eta_2$, we place end-of-the-world (EOW) branes, denoted by $Q^{h(1)}$ and $Q^{h(2)}$, on which we impose Neumann boundary conditions
\ba
K_{ab}-h_{ab}K=-Th_{ab},\label{NBY}
\ea
where $K_{ab}$ is the extrinsic curvature (with outward-pointing normal $n^a$) and $T$ denotes the brane tension. The Neumann boundary conditions (NBC) are satisfied provided the tension of the branes are
\ba
T^{h(i)}=\frac{d-1}{d}K^{h(i)}=\frac{d-1}{\eta_{i}}, \label{NBYa}
\ea
with $i=1,2$ being labels of two EOW branes. By analogy with wedge holography in AdS \cite{Akal:2020wfl}, authors proposed that ``{\it gravity in the $(d+1)$-dimensional wedge region $W^h$ (\ref{etawdge}) is dual to a $(d-1)$-dimensional CFT living on the sphere $S^{d-1}$ at the asymptotic tip $\rho\to \infty$}''. A cutoff $\rho=\rho_\infty$ as in (\ref{cutoffb}) regulates the boundary theory. The authors provided evidence for this duality via the explicit computations of the partition function, holographic entanglement entropy, and scalar perturbations.

Each hyperbolic slice $H^d$ at fixed $\eta$ preserves an $SO(1,d)$ symmetry, corresponding to the Lorentz group of the parent $(d+1)$-dimensional Minkowski spacetime. This matches the conformal symmetry of the Euclidean CFT on $S^{d-1}$. In the special case $d=3$, the symmetry enhances to two copies of the Virasoro algebra, originating from the superrotation symmetry in $\mathbb{R}^{1,3}$ which is naturally identified with the conformal symmetry of the dual two-dimensional CFT.

The analysis further done by the authors indicates that the dual CFT on $S^{d-1}$ is non-unitary. This feature is expected: the interval in the time-like direction (\ref{etawdge}), orthogonal to the hyperbolic slices $H^d$ has been included as an internal dimension, even though each slice separately admits a standard AdS/CFT interpretation. The situation is closely analogous to dS/CFT, where the dual CFT is likewise non-unitary, as inferred from central charge arguments \cite{Maldacena:2002vr} and supported by explicit realizations \cite{Anninos:2011ui,Cotler:2019nbi,Hikida:2021ese,Hikida:2022ltr,Dey:2024zjx}.

\subsubsection{Wedge holography for de Sitter slices} \label{sec:dS}
%%%%%%%%%%%%%%%%%%%%%%%%%%%%%%%%%%%%%%%%%%%%%%%%
As a second example of flat space wedge holography, consider the $(d+1)$-dimensional wedge $W^{ds}$, defined by restricting the de Sitter slicing (\ref{dspatch}) to the region
\ba
r_1\leq r\leq r_2,  \label{etawdgee},
\ea
as illustrated in the right panel of Fig.~\ref{fig:wedgeah}. The two boundaries at $r=r_1$ and $r=r_2$ are taken to be EOW branes, denoted $Q^{ds(1)}$ and $Q^{ds(2)}$, where we impose the Neumann boundary condition (\ref{NBY}). Solving this condition yields the brane tensions 
\ba
T^{ds(i)}=\frac{d-1}{d}K^{ds(i)}=\frac{d-1}{r_{i}}, \label{NBYbaa}
\ea
where $i=1,2$ denotes two end-of-the-world branes.

The authors proposed that ``{\it gravity on the wedge $W^{ds}$(\ref{etawdgee}) is holographically dual to a $(d-1)$-dimensional CFT defined on a sphere $S^{d-1}$}. Although the wedge has two asymptotic boundaries located at $t=-\infty$ and $t=\infty$, they are identified through the antipodal map. A cutoff $t=\pm t_\infty$, analogous to (\ref{cutoffb}), regulates the theory. 
As in the hyperbolic case, each de Sitter slice $dS_d$ at a fixed value of radial coordinate exhibits an $SO(1,d)$ symmetry, corresponding to the Lorentz symmetry of the ambient $(d+1)$-dimensional Minkowski spacetime, which in turn matches the conformal symmetry of the Euclidean CFT on $S^{d-1}$. 
In the case $d=3$, this symmetry is enhanced to two copies of the Virasoro algebra, arising from the superrotation symmetry of $\mathbb{R}^{1,3}$ \cite{Barnich:2009se,Barnich:2010ojg}, and is naturally identified with the conformal symmetry of a dual two-dimensional CFT.

This setup can be viewed as a de Sitter analogue of wedge holography in AdS \cite{Akal:2020wfl}, with the wedge defined by extending a $dS_d$ slice along a spatial width. Consequently, the dual CFT on $S^{d-1}$ is again expected to be non-unitary, in close analogy with the dS/CFT correspondence \cite{Strominger:2001pn,Maldacena:2002vr,Anninos:2011ui,Cotler:2019nbi,Hikida:2021ese,Hikida:2022ltr,Dey:2024zjx}. Authors provided support to this duality via computing the partition function, holographic entanglement entropy, and scalar field perturbation.

\section*{Acknowledgements}
I would like to thank the organizers of ST$^4$ for their hospitality. I am grateful to Alok Laddha for insightful comments and to Ghanshyam Date for helpful suggestions, which have greatly improved these lecture notes. I would like to thank Krishna Jalan, Kanhu Kishore Nanda, Somnath Porey, and Hemant Rathi for useful discussions on various topics during the lectures. I would also like to thank my collaborators and other researchers [I have learned extensively from them] with whom I have had discussions. This work is partially supported by a grant to CMI from the Infosys Foundation.

\appendix
\section{Excercises}
In this section, we list all the excercises given during the lectures at one place. We provide solution of one problem.
\begin{itemize}
\item {\bf E1}: Derive the generator for infinitesimal translation (\ref{ict-a}). 

\item {{\bf E2}: Derive the generators of infinitesimal Lorentz rotations (\ref{ict-a-x-as}).}

\item {\bf E3}: Derive the generator of infinitesimal special conformal transformations (\ref{SCTD}). 

\item  {\bf E4}: Prove that the finite special conformal transformation and the associated scale factor with it is given as follows
\begin{eqnarray}\label{eq:eq3p20-ex}
& & x'^\mu = \frac{x^\mu-(x\cdot x) b^\mu}{1-2(b\cdot x)+(b\cdot b)(x \cdot x)}, \nonumber \\
& &  \Lambda(x) = \left( 1-2(b\cdot x) + (b\cdot b)(x \cdot x)  \right)^2.
\end{eqnarray}

\item  {\bf E5}: Use $x'^\mu$ from (\ref{eq:eq3p20}) and prove that
\begin{equation}
\frac{x'^\mu}{x' \cdot x'} = \frac{x^\mu}{x \cdot x} - b^\mu.
\label{eq:eq2p71-ex}
\end{equation}

\item {\bf E6}: For the generatores (\ref{Jgen}), prove the following algebra
\begin{equation}
\left[ J_{mn},J_{pq}   \right] = i \left( \eta_{mq}J_{np} + \eta_{np}J_{mq} - \eta_{mp}J_{nq} - \eta_{nq}J_{mp} \right). \label{eq:129-ex}
\end{equation}

\item {\bf E7}: Derive the following algebra associated with generators $l_n= - z^{n+1} \partial$ and $\bar{l}_n= - \bar{z}^{n+1} \bar{\partial}$.
\begin{align} \label{W-A-ex}
[\ell_m, \ell_n] &=  (m-n)\ell_{m+n}, \nonumber \\
[\bar{\ell}_m, \bar{\ell}_n] &= (m-n)\bar{\ell}_{m+n} , \\
[\ell_m, \bar{\ell}_n] &= 0. \nonumber
\end{align}

\item {{\bf E8}: Derive the form of brown york stress tensor as given in \cite{Balasubramanian:1999re}. Use it to obtain the stress tensor for $AdS_3$ background and, from the trace, derive the central charge. Finally, show that the stress tensor is traceless in $AdS_4$ spacetime.}

\item {\bf E9}: Compute the pseudo entropy in $dS_3/CFT_2$ correspondence by following \cite{Narayan:2023zen}.

\item  {\bf E10}: Derive the generator associated with scale transformations (\ref{ict-a-x-s}).\\
{\bf Solution}: Generic infinitesimal transformations may be written as
\begin{eqnarray}
x'^\mu &=& x^\mu + \epsilon_a \frac{\delta x^\mu}{\delta\epsilon_a} \nonumber \\
 \phi'(x') &=& \phi (x) + \epsilon_a \frac{\delta \phi(x)}{\delta \epsilon_a(x)}, \label{eq:inftrans-ex}
\end{eqnarray}
%where $F$ is the function relating the new field $\phi'$ evaluated at the transformed coordinate $x'$ to the old field $\phi$ at $x$
%$$
%\phi'(x')=F(\phi(x)),
%$$
%and we are keeping infinitesimal parameters $\{ \epsilon_a \}$ to first order. 
The convention we follow is that the generator $G_a$ of a transformation is given by
\begin{equation}
\phi'(x)-\phi(x) \equiv i \epsilon_a G_a \phi(x),
\end{equation}
so that 
\begin{equation}
i G_a \phi(x) =\frac{\delta \phi(x)}{\delta \epsilon_a(x)}= \frac{\delta x^\mu}{\delta \epsilon_a} \partial_\mu \phi(x).  \label{eq:eq1p20}
\end{equation}
Under infinitesimal scale transformations with generator $D$, $x^\mu$ transform as: $x^\mu \rightarrow e^\epsilon x^\mu \approx (1+\epsilon)x^\mu$ so that
\begin{eqnarray}
i D \phi(x) &=&  \frac{\delta x^\mu}{\delta \epsilon}\partial_\mu \phi(x) .
\end{eqnarray}
The above equation implies that $ D = -i x^\mu \partial_\mu$ this is what we want.

\end{itemize}

\end{document}